\documentclass[twocolumn,superscriptaddress, aps,pre ]{revtex4-1}

\usepackage{color}
\usepackage{graphicx}
\usepackage{hyperref}
\usepackage{amsmath,amssymb}
\usepackage{epstopdf}
\usepackage{subcaption}
\graphicspath{{figs/}}
\begin{document}
\title{Sandpile on uncorrelated site-diluted percolation lattice;\\
 From three to two dimensions}
 
\author{M. N. Najafi}
\affiliation{Department of Physics, University of Mohaghegh Ardabili, P.O. Box 179, Ardabil, Iran}
\email{morteza.nattagh@gmail.com}
\author{H. Dashti-Naserabadi}
\affiliation{Physics and Accelerators Research School, NSTRI, AEOI 11365-3486, Tehran,Iran}
\email{h.dashti82@gmail.com}

\date{\today}

\begin{abstract}
The BTW sandpile model is considered on three dimensional percolation lattice which is tunned with the occupation parameter $p$. Along with the three-dimensional avalanches, we study the energy propagation in two-dimensional cross-sections. We use the moment analysis to extract the exponents for two separate cases: the critical ($p=p_c\equiv p_c^{3D}$) and the off-critical ($p_c<p\leq 1$) cases. The three-dimensional avalanches at $p=p_c$ has exponents like the regular 2D BTW model, whereas the exponents for the 2D cross-sections have serious similarities with the 2D critical Ising model. The moment analysis show that finite size scaling theory is the fulfilled, and some hyper-scaling relations are confirmed. For the off-critical lattice, the exponents change logarithmically with $p-p_c$, for which the cut-off exponents $\nu$ drop discontinuously from $p=p_c$ to the other values. The analysis for the 2D cross-sections show a singular behavior at some $p_0\approx p_c^{2D}$ ($p_c^{3D}$ and $p_c^{2D}$ being three- and two-dimensional percolation thresholds). We argue that there are two separate phases in the cross-sections, namely $p_c^{3D}\leq p<p_c^{2D}$ which, due to lack of 2D percolation cluster, has no thermodynamic limit, and $p\geq p_c^{2D}$ having the chance to involve percolated clusters.
\end{abstract}

\pacs{05.40.-a, 45.70.Cc, 11.25.Hf, 05.45.Df}
\keywords{3D BTW, Cross-Section, Percolation Lattice}

\maketitle

\section{Introduction}
There are theoretical and experimental interests in the notion of critical phenomena on the fractal lattices as a long-standing problem in physics. Examples of experimental motivations are the voids of percolating clusters which are filled by (commonly magnetite) nano-particles of a ferromagnetic fluid \cite{kose2009label,kikura2004thermal,matsuzaki2004real,philip2007enhancement,kim2008magnetic,benkoski2008self,kikura2007cluster}. In the theoretical side the main contribution was made by Gefen, \textit{et. al.} \cite{gefen1980critical} in which it was claimed that the critical behavior of the models on the fractal geometries (for which no lower critical dimension can be defined) is tuned by the detail of the topological quantities of the fractal lattice. The cluster fractal dimension, the order of ramification and the connectivity are some examples of these quantities \cite{gefen1980critical}. This mixing of two statistical models (one as the dynamical model and the other as the host for the first one) may also be interpreted as the interplay between two statistical models from which some new non-trivial critical behaviors can emerge. Examples are the Ising model in a BTW sandpile \cite{koza2007ising} and the Ising model on the percolation lattices \cite{najafi2016monte,cambier1986distribution,coniglio1989fractal,coniglio1980clusters,wang1990fractal,saberi2009thermal,davatolhagh2012critical,scholten1997monte}. The fluid movement in the porous media is the other important example which can be modeled by invasion percolation \cite{wilkinson1983invasion}, or directly by the Darcy's reservoir model on the percolation lattices \cite{najafi2015geometrical,najafi2016water}. The later application contains the important concept of self-organized criticality on the percolation lattices \cite{najafi2015geometrical,najafi2016bak}. Due to its important aspects, this subject has been investigated in many papers. However a comprehensive understanding of this subject is missing yet in the literature, e.g. there is no overall agreement on their exact universality classes. \\
Sandpile model on the fractal (and other) lattices is an important topic in the literature which has theoretical \cite{najafi2016bak,dhar1990abelian} and empirical \cite{najafi2016water,daerden2001waves} attractions. Fortunately the critical behaviors of the sandpile model in three and two dimensions have been vastly studied analytically \cite{Bak1987Self,Dhar1990Self,Dhar1999Abelian,majumdar1992equivalence,najafi2013left,Azimi2011Continuous,Majumdar1991Height,Majumdar1992Exact,Manna1990Cascades} and numerically \cite{Lubeck1997BTW,Ktitarev2000Scaling,najafi2013numerical,asasi2015continuous,Lubeck1997Numerical,Moghimi2005Abelian,najafi2012avalanche,najafi2014bak,najafi2016bak,najafi2016water,najafi2012observation}. Despite of this huge literature, a little attention has been paid to  the general aspects of the sandpile model on the percolation lattices, e.g. the dependence on the dimensionality. In this respect the exact determination of the exponents is a challenging problem \cite{asasi2015continuous}, and a detailed finite-size analysis is required. Along with these issues, the problem of energy propagation in a subset of the original system, e.g. the propagation of avalanches in two-dimensional cross-sections of the three-dimensional system, has especial importance \cite{dashti2015statistical}. In the theoretical side such an investigation reveals which model lives in a $d-1$ dimensional subsystem of a $d$ dimensional system~\cite{dashti2017bak}.\\
In this paper we focus on the critical properties of the BTW model on the three-dimensional site-diluted cubic percolation lattice in terms of the parameter $p$ which tunes the occupation probability of the lattice. The paper is divided to two distinct parts: In the first part we consider $p$ to be the critical one, i.e. $p=p_c=\text{percolation threshold}$. We name this as the \textit{critical regime}. In the second part we analyze the case $p_c<p\leq 1$ which is named as the \textit{off-critical regime} (note that it does not imply that the system is not critical, but it displays the off-criticality of the percolation lattice). The moment analysis, as a precise tool for extracting the exponents is employed. We show that in the latter case the exponents show some non-trivial logarithmic features having its root at the properties of the percolation clusters.\\
Parallel to the three-dimensional analysis, the energy propagation in two-dimensional cross-sections is also investigated and some exponents emerge which respect to some hyper-scaling relation for the fractal dimensions. The fractal dimension of loops (the exterior perimeter of the connected avalanches) at $p=p_c$ is determined to be $D_F^{ \text{cross-sections}}\approx 1.37$ compatible with the fractal dimension of loops of the spin clusters of the Ising model, i.e. $D_F^{\text{2D-Ising}}=\frac{11}{8}$ \cite{najafi2015observation}, the result which is compatible with the sandpile model on two-dimensional site-diluted square percolation lattice \cite{dashti2015statistical}. The other exponents also support this hypothesis. In the off critical regime we observe a singular behavior, having its root in the non-percolating character of 2D percolation systems for $p_c^{3D}<p<p_c^{2D}$, in which $p_c^{3D}$ and $p_c^{2D}$ are the percolation thresholds in $3D$ and $2D$ systems.\\
The paper has been organized as follows: In the SEC.\ref{Motivation} we introduce the problem and its motivations. Section \ref{critical} has been devoted to the statistical properties of the model in the critical regime. The off-critical exponents have been analyzed in SEC. \ref{off-critical} with two subsections: three and two dimensional systems. We end the paper by a conclusion in SEC. \ref{conclusion}. 
\section{Motivation and Model definition}\label{Motivation}
\begin{figure*}
\begin{subfigure}{0.45\textwidth}\includegraphics[width=\textwidth]{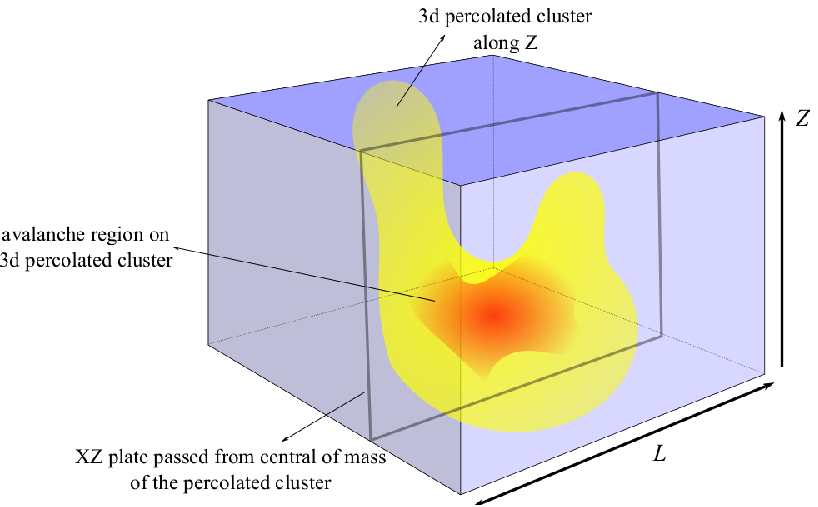}
\caption{}
\label{fig:3d-avalanche}
\end{subfigure}
\begin{subfigure}{0.30\textwidth}\includegraphics[width=\textwidth]{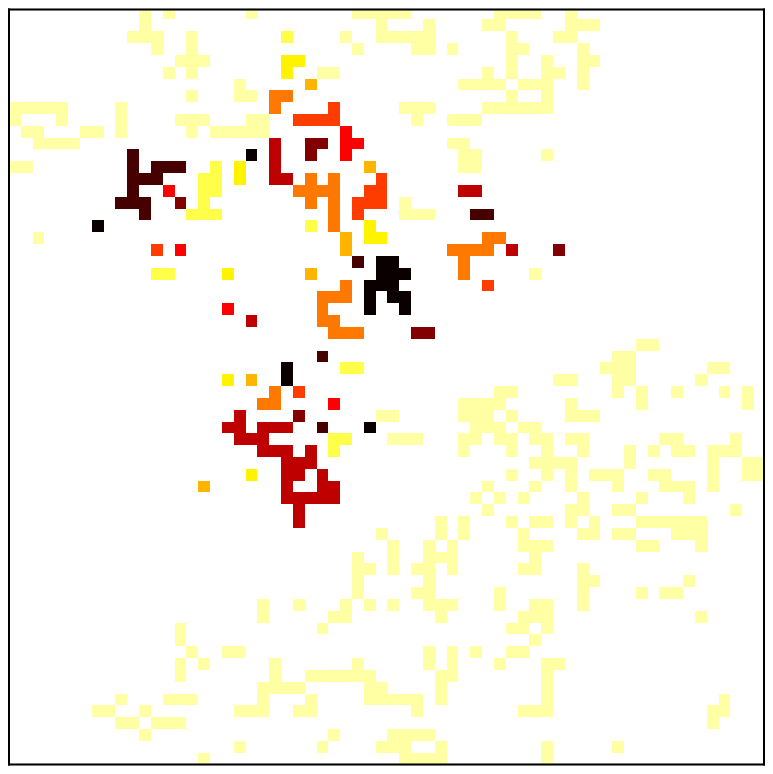}
\caption{}
\label{fig:2Dsample}
\end{subfigure}
\caption{(Color Online) (a). A schematic set-up of the problem. The 2D cross-section has been shown right at $y=\frac{L}{3}$. The resulting 2D avalanche (in the cross-section) can apparently be non-simply-connected (this is true also for 3D avalanches for the presence of non-active sites, but for the cross-sections it can be due to the distortions of the 3D avalanche). (b) A sample of 2D height configuration in the cross-sections.}
\label{fig:samples}
\end{figure*}
Let us first introduce the ordinary BTW model on the regular cubic lattice which defines the dynamics of sand grains. The sand grains are distributed randomly through the lattice, so that we have a local height filed $h$ over the lattice, for which the constraint is that no site has the height higher than $2d$ ($d=$ spatial dimension of the system which is three in this paper), i.e. $h(i)$ takes the numbers from the set $\lbrace{1, 2, ... , 2d}\rbrace$ for each site $i$. The system is open, i.e. adding or losing energy is allowed. The dynamic of the system is as follows: A random site ($i$) is chosen and a grain is added to this site, i.e. $h(i)\rightarrow h(i)+1$. If the resulting height is lower than a critical value ($h(i)\leq h_c=2d$), another site is chosen for adding a grain. But if this height exceeds the critical value ($h(i)>h_c$), then this site becomes unstable and topples. During this toppling, the height of the original site $i$ is lowered by a number equal to its neighbors ($h(i)\rightarrow h(i)-2d$) and the content of each of its neighbors is increased by one in such a way that the total number of grains is conserved. The single toppling process can be expressed via the relation $h(i)\rightarrow h(i)-\Delta_{i,j}$ in which
\begin{equation}
\Delta_{i,j}=
\begin{cases} -1 & i \text{ and } j \text{ are neighbors} \\
2d & i=j \\ 
0 & \text{other}
\end{cases}
\end{equation}
As a result of this toppling, the neighboring sites may become unstable and topple. This process continues until reaching the state in which all sites of the system become stable. An avalanche is defined as the chain of activity that is triggered by adding a grain to a stable state until another stable state is reached. Now another site is chosen for injection and the process continues. Generally we have two kinds of configurations: transient and recurrent ones. The transient configurations may happen once in the early evolution steps and shall not happen again and the recurrent configurations take place in the steady state of the system. In this state of the system, the energy input and output of the system is statistically equal and the statistical observables of the system are statistically constant. All of the configurations in this state occur with the same probability. For a good review see~\cite{Dhar1999Abelian}. The fact that this system organizes itself in the critical state, has been intensively argued by many researchers \cite{vespignani1998driving,rossi2000universality,karmakar2004directed}.\\
Now we turn to the main problem of the present paper, i.e. the BTW model on the site-diluted cubic percolation lattice. A percolation lattice is constructed simply by the following rule: each lattice site is occupied (active) by the probability $p$ and is un-occupied (inactive) by the probability $1-p$. There is a critical occupation probability (percolation threshold) $p_c$ such that for $p\geq p_c$ there are some connected clusters (involving the set of sites of the same type) which percolates, namely percolated clusters. For the cubic lattice the critical threshold is nearly $p_c\approx 0.32$. By the critical model we mean the BTW model on the lattice with $p=p_c$. For other occupations ($p_c<p\leq 1$) we use the phrase \textit{out of critical} model, although the model shows critical properties. At each $p$ a single spanning cluster is chosen as the host for the sandpile model which should contact the boundaries (note that the cluster should contain boundary sites to dissipate sand grains). For the simulation of the grain propagation in the percolated clusters we use the simple rules of the BTW model as stated in the previous section, except the fact that the sand grains cannot enter the un-occupied sites, i.e. when an unstable site has $z$ occupied neighbors, during a local toppling its grain content decreases by $z$ and each of its activate neighbors increases by one. Also the sand grains have the chance to leave the system via the boundary sites of the percolated cluster. To have a true statistical analysis on both BTW dynamics and percolation problem, we force the spanning cluster change after extracting some avalanche samples. Therefore our statistical analysis contains averaging over BTW as well as percolation configurations. \\
The problem of two-dimensional propagation of sand grains (energy) in three dimensional systems seems to be very important from both theoretical and experimental sides. More precisely the important question in the theoretical physics is that how the information in $d+1$ dimensions would be reflected to its $d$ dimensional subsystem. For this purpose one should map the original $d+1$ dimensional model to a $d$-dimensional one and measure how some information are lost and how the degrees of freedom in the subtracted dimension affect the $d$-dimensional model, i.e. which model lives in the lower dimensional system. If the subtracted dimension be temporal, then one is looking at a \textit{frozen} model with no dynamics. The investigation of the contour lines of statistical systems~\cite{najafi2012observation} and the ground state of the quantum systems~\cite{najafi2016scale} are some examples. A more interesting situation is the case in which the subtracted dimension is spatial one, like the holography principle. The example is the cross sections of three-dimensional BTW model which is proposed to share some critical behaviors as the 2D Ising model~\cite{dashti2015statistical,dashti2017bak}. In this paper along with the three-dimensional analysis, we study the energy propagation through two-dimensional slices, i.e. cross-sections of the three-dimensional system. A schematic graph is presented in Fig. \ref{fig:3d-avalanche} showing the total set up of a percolated host cluster with a three-dimensional avalanche and its cross-section. Also a real two-dimensional (2D) sample is presented in the Fig. \ref{fig:2Dsample}, to visualize how the height-field varies over the 2D sample. In Fig. \ref{fig:3d-avalanche}, a percolated cluster along the $Z$ axis has been shown with an avalanche which has been shown by a red area. A $X$-$Z$ plate has been sketched in this figure, showing the mentioned cross-sections which passes from the center of mass of the host cluster. In the Fig. \ref{fig:2Dsample} the set of color sites show the cross section of the percolated cluster. The light yellow sites are not contained in the avalanche and the other colors show various connected components of an avalanche cluster in that cross-section.

In the critical state one expects a power-law behavior for the local and geometrical quantities. For example for the ordinary BTW model the distribution functions behave like $P(x)\sim x^{-\tau_x}$ ($x=$ the statistical quantities in three- and two-dimensional systems). The exact determination of these exponents plays a vital role and a detailed finite-size analysis is required. For finite systems, the finite-size scaling (FSS) theory predicts that \cite{Lubeck1997BTW}:
\begin{equation}
P_x(x,L)=L^{-\beta_x}g_x(xL^{-\nu_x}),
\label{eq:FSS}
\end{equation}
in which $g$ is a universal function and $\beta_x$ and $\nu_x$ are the exponents corresponding to $x$. A simple dimensional analysis shows that $\tau_x=\frac{\beta_x}{\nu_x}$, which will be tested for all observables in this paper. 
The exponent $\nu_x$ determines the cutoff behavior of the probability distribution function. If FSS works, all distributions $ P_x (x,L) $ for various system sizes have to collapse, including their cutoffs. Then the argument of the universal function $ g_x $ has to be constant. One can simply show that $r_{\text{cutoff}}\sim L^{\nu_r}$, i.e., the cutoff radius should scale linearly with the system size $ L $ ($\nu_r=1$), so that for all observables one gets $ \nu_x = \gamma_{xr} $ \cite{Lubeck1997BTW}.\\
The mono-fractalinty and multi-fractality of the sandpile models is the notion which is served as an important issue in the literature. Before closing the section, we mention some points on the multi-fractal structure of the model which is a long-standing debate in the literature. In fact the relation \ref{eq:FSS} is only correct for mono-fractal systems. To investigate this, we use the method of moment analysis presented in \cite{tebaldi1999multifractal}. To this end, we should calculate the $q$th moment of the $x$ variable $\left\langle x^q\right\rangle$ ($x=$ the statistical observable in each dimension), defined by:
\begin{equation}\label{eq:moment}
\left\langle x^q\right\rangle_L=\int P_x(x,L)x^qdx\sim L^{\sigma_x(q)},
\end{equation}
in which $\sigma_x(q)=\nu_x\left(q-\tau_x+1\right)$ for mono-fractal systems. It is seen that for mono-fractal systems $\sigma_x(q)$ has the linear behavior in terms of $q$, i.e. $\sigma_x(q+1)-\sigma_x(q)=\nu_x$. It is a serious test for mono-fractality and multi-fractality of the system. In addition the exponents can be extracted from this analysis. In the following sections we use this analysis. Also we note that there is a hyper-scaling relation between the $\tau$ exponents and the fractal dimensions $\gamma_{x,y}$, which are defined by the relation $x\sim y^{\gamma_{x,y}}$, namely:
\begin{equation}
\gamma_{x,y}=\frac{\tau_y-1}{\tau_x-1}.
\end{equation}
This relation is valid only when the conditional probability function $p(x|y)$ is a function with a very narrow peak for both $x$ and $y$ variables.

\section{Critical case; $p=p_c$}
\label{critical}

\begin{figure*}
\begin{subfigure}{0.45\textwidth}\includegraphics[width=\textwidth]{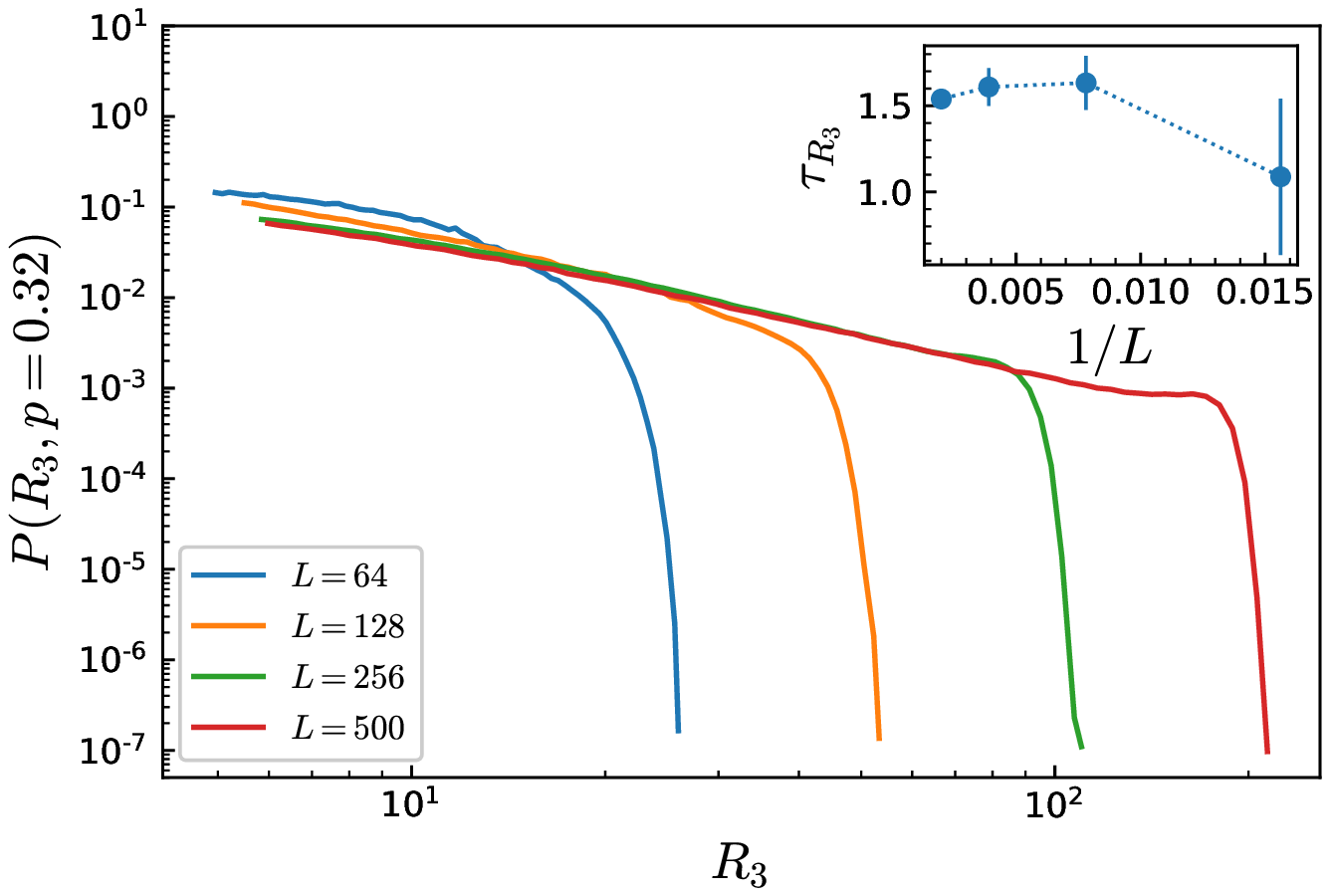}
\caption{}
\label{fig:P_R3_pc}
\end{subfigure}
\begin{subfigure}{0.45\textwidth}\includegraphics[width=\textwidth]{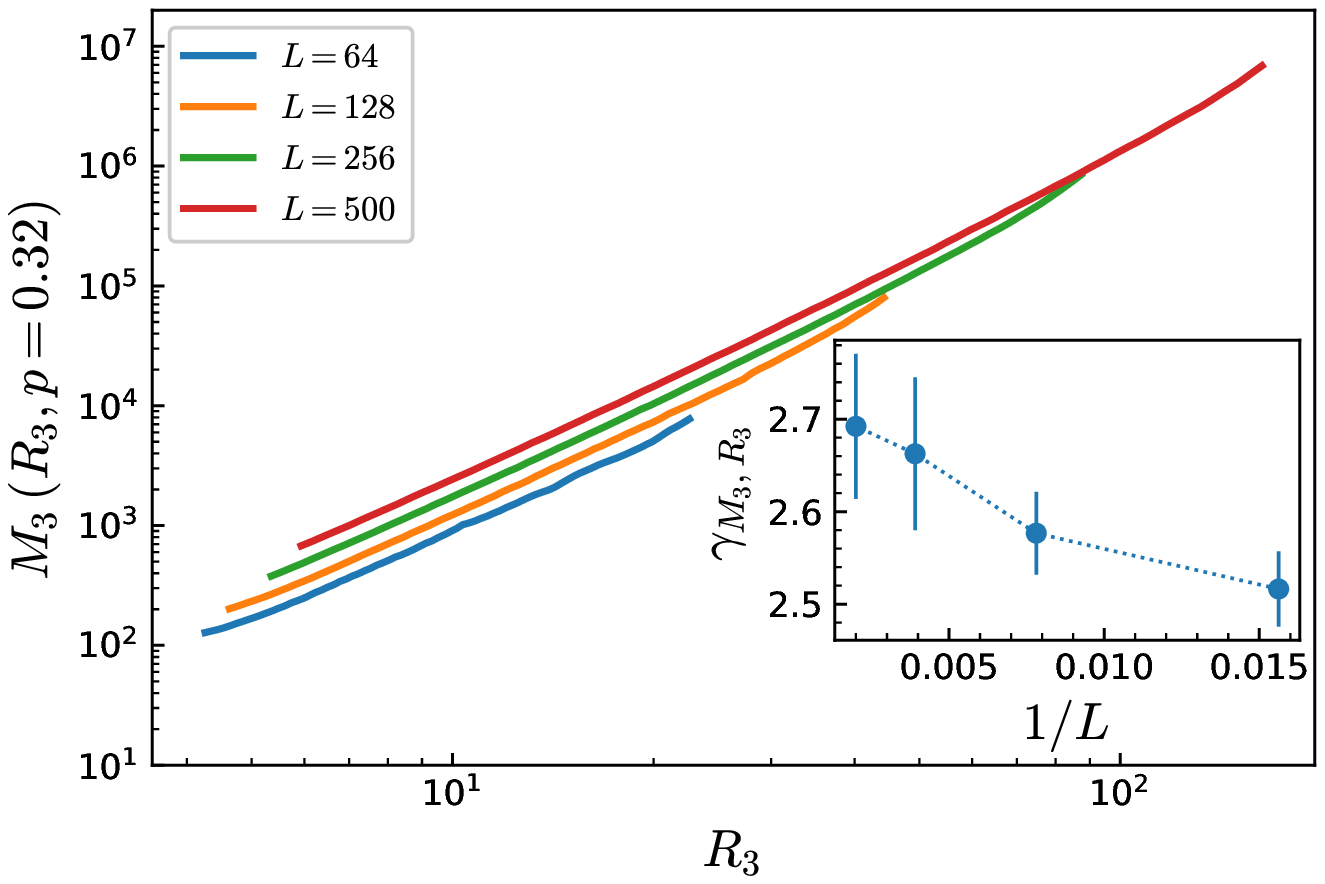}
\caption{}
\label{fig:M3_R3_pc}
\end{subfigure}
\begin{subfigure}{0.45\textwidth}\includegraphics[width=\textwidth]{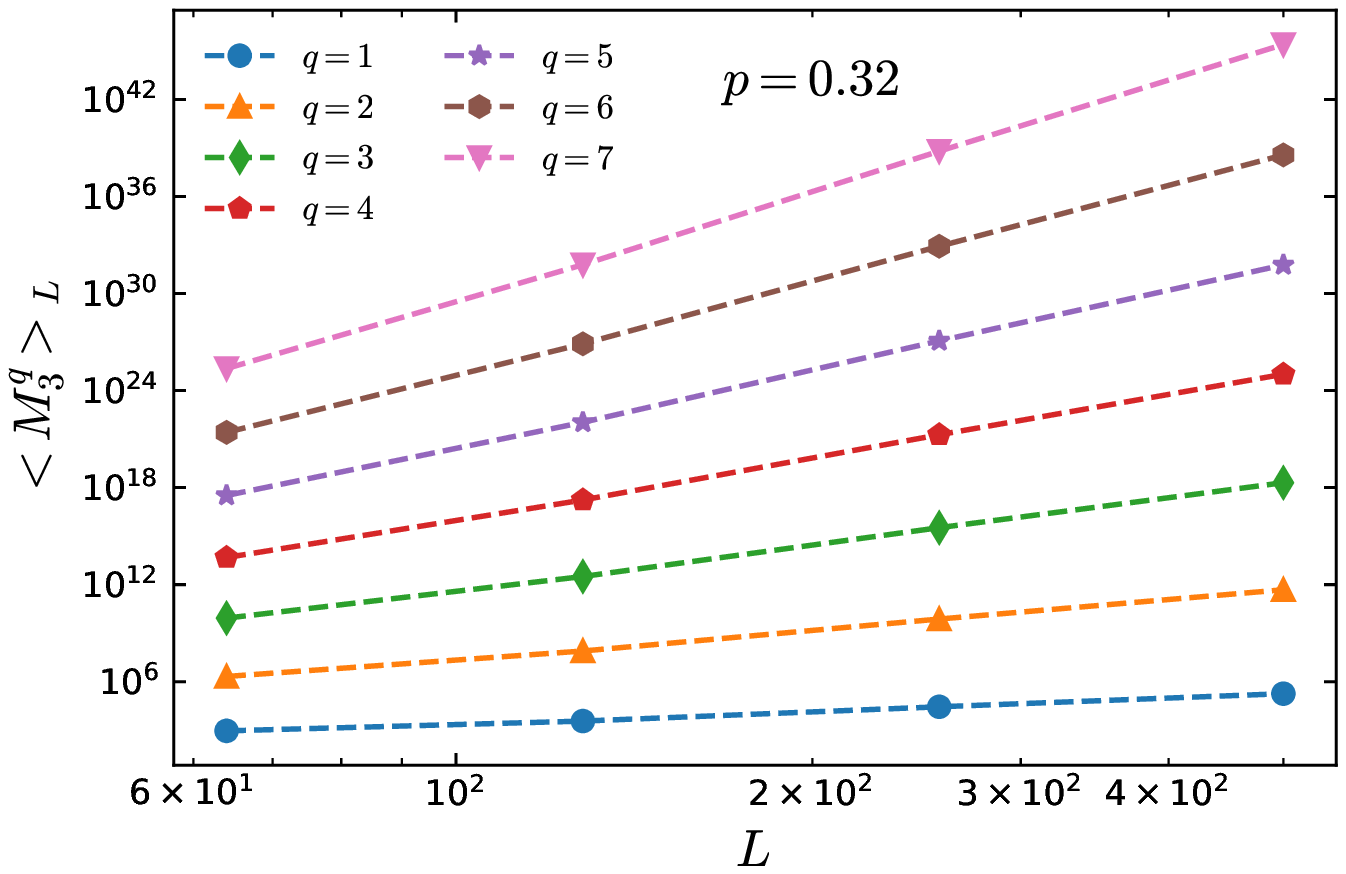}
	\caption{}
	\label{fig:M3_q}
\end{subfigure}
\begin{subfigure}{0.45\textwidth}\includegraphics[width=\textwidth]{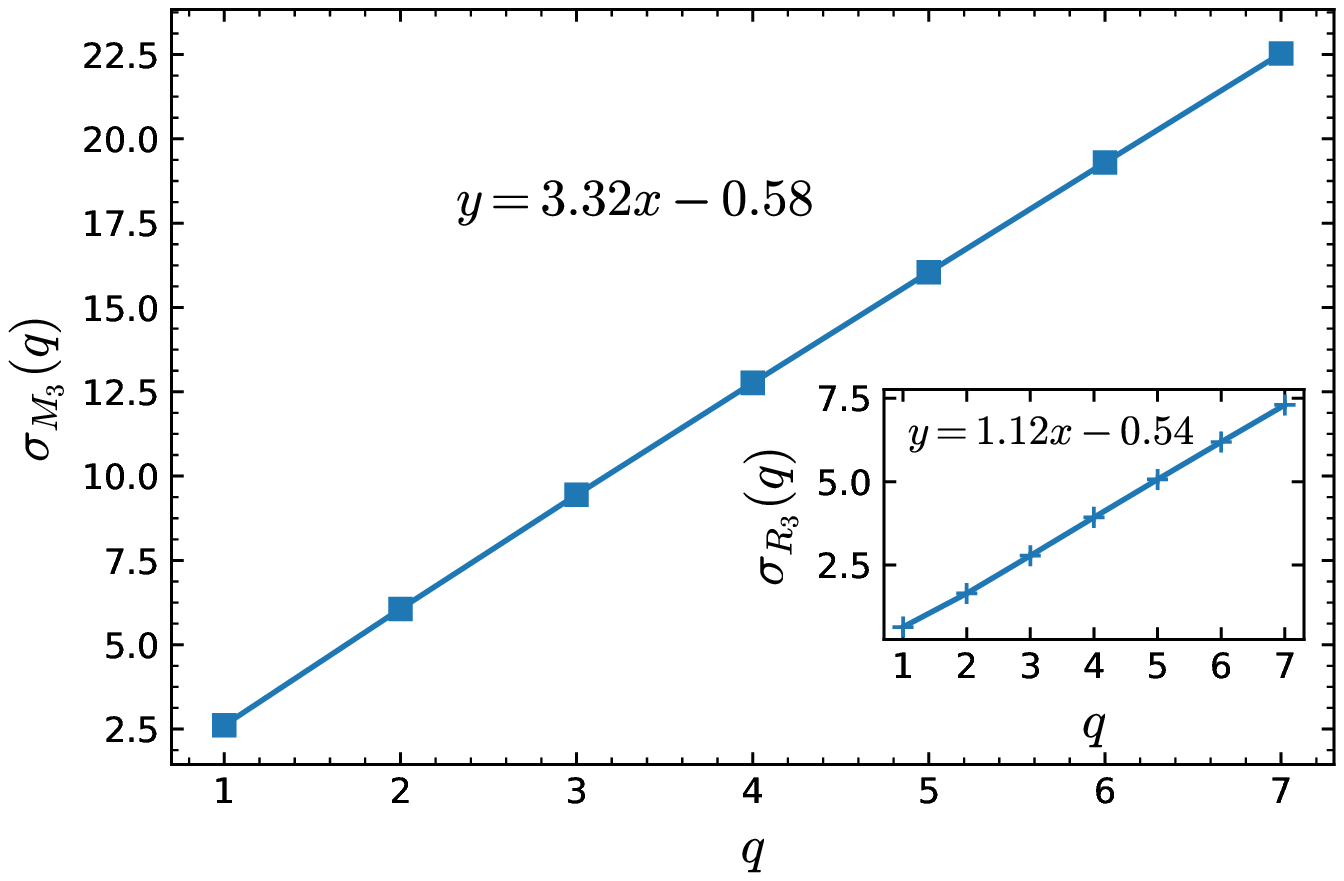}
	\caption{}
	\label{fig:sigma_MR_pc}
\end{subfigure}
\caption{(Color Online) The exponents of the critical case $p=p_c$ for the 3D system. (a) The distribution function of the three-dimensional gyration radius ($R_3$). Inset: the finite size dependence of $\tau_{R_3}$. (b) The finite-size dependence of the fractal dimension $\gamma_{M_3R_3}$ defined as $ M_3 \sim  R_3^{\gamma_{M_3R_3}} $. (c) The $q$th moment of $M_3$ in terms of $L$ for various rates of $q$. (d) The exponent $\sigma_{M_3}(q)$ and $\sigma_{R_3}(q)$ in terms of $q$. When the graph is fitted by $\sigma_x=\nu_x q+b_x$, then the corresponding $\tau$ exponent is obtained via $\tau_x=1+\frac{b_x}{\nu_x}$. }
\label{fig:p_c}
\end{figure*}
In this section we focus on the case $p=p_c$. The analysis in the critical occupation is very important since, in the thermodynamic limit, only at this occupation value ($p_c$) the system has the chance of being distinct from the $p=1$ case. This is because of the fact that in the thermodynamic limit, all $(p_c<p\leq 1)$-percolation systems are identical to the $(p=1)$-percolation system and all of their behaviors are the same as $p=1$ one \cite{najafi2016bak,najafi2016water}. Therefore one expects that the size-dependent exponents be such that for $L\rightarrow \infty$ only the exponents for $p=p_c$ are distinct from the $p=1$ case.\\
We have two types of quantities in 3D systems and their 2D cross sections: the fractal dimensions and the distribution functions for the statistical observables. The three dimensional quantities studied in this paper are as follows:\\
- The avalanche mass $(M_3)$ which is the total number of sites involved in a three-dimensional avalanche\\
- The three-dimensional gyration radius $R_3$ which is defined by $R_3^2\equiv \frac{1}{M_3}\sum_{i=1}^{M_3} ( \vec{r}_i-\vec{r}_{\text{com}} )^2$, i.e. it is the gyration radius of points involved in a three-dimensional avalanche. In this formula $\vec{r}_i\equiv(x_i,y_i,z_i)$ is the position vector of the $i$th point of the avalanche and $\vec{r}_{\text{com}}\equiv (x_{\text{com}},y_{\text{com}},z_{\text{com}})\equiv \frac{1}{M_3}\sum_{i=1}^{M_3}\vec{r}_i$ is the center of mass of the avalanche.\\
- The number of topplings in a three-dimensional avalanche $s_3$.\\
Let us first consider the 3D avalanches. The numerical analysis right at $p=p_c$ is very hard and bothersome, since the typical time needed to reach a stable configuration is related inversely to the number of the boundary sites which is very low in $p=p_c$. On the other hand due to the diluteness of sites the samples are commonly small, the fact which affects the quality of the results. The way out of these deficiencies is to firstly choose the percolated cluster with larger boundary sites, and secondly increase the number of samples, so that some (rare) large samples have the chance to appear. In this work we have generated over $5\times 10^6$ samples for each $L$ and $p$. The simulation on each SPC$_p$ (spanning percolation cluster tuned by the occupation parameter $p$) has been started with a random $h$ configuration and the statistical analysis has been carried out in the steady states. The lattice sizes considered in this work are $L=64,128,256$ and $500$. After extracting $10^3$ avalanche samples, another percolated cluster is chosen and so on. The SPC's have been chosen so that the fraction $\frac{\text{N.O. boundary sites}}{\text{N.O. bulk sites}}$ is not that small to prevent the average time duration of avalanches of becoming very high. For the distribution functions there are some cut values above which the linear behavior of the log-log graph is destroyed and the graph falls off rapidly. In this scale the finite size plays the dominant role. This cut-off for the observable $x$ scales with $L$ by the relation $x^{\text{cut}}\sim L^{\nu_x}$ in which $\nu_x$ has been defined in the previous section. This exponent can be obtained by means of moment analysis. We have observed that all of the fractal dimensions at $p=p_c$ (three and two dimensional systems) scale with $L$ by the relation:
\begin{equation}
\gamma_{xy}^{p_c}(L\rightarrow\infty)-\gamma_{xy}(L)^{p_c}\propto \frac{1}{L}
\end{equation}
in which the quantities $\gamma_{xy}(L\rightarrow\infty)^{p_c}$ are of interest. The $\tau$ exponents however saturate in some final value $\tau_x(L\rightarrow L_{\text{max}})^{p=p_c}$. This should be compared with the one for two dimensional regular BTW model, i.e. $\tau_x(L)^{\text{regular BTW}}_{d=2}=\tau_{x,\infty}-\frac{\text{const.}}{\ln(L)}$ \cite{Lubeck1997Numerical}. \\
In the Fig. \ref{fig:P_R3_pc} we have shown the distribution function of the three-dimensional gyration radius $P(R_3)$ for various rates of the lattice sizes $L$. As can be seen in the inset of Fig.~\ref{fig:P_R3_pc}, $\tau_{R_3}$ saturates at final value $1.5\pm 0.1$. The results of the moment analysis have been shown in Figs.~\ref{fig:M3_q} and~\ref{fig:sigma_MR_pc}. The linearity of the log-log plot of $\left\langle M_3^q\right\rangle$ in terms of $L$ is evident in the Fig.~\ref{fig:M3_q} whose slopes increase as $q$ increases. The corresponding exponent $\sigma_{M_3}(q)$ and $\sigma_{R_3}(q)$ has been shown in Fig.~\ref{fig:sigma_MR_pc}, from which one finds $\nu_{M_3}=3.32(4)$, $\nu_{R_3}=1.12(4)$, $\tau_{M_3}=1.18(4)$ and $\tau_{R_3}=1.49(4)$. The fact that $\nu_{R_3}$ is near the unity is expected for the critical systems. It is seen that $\nu_{M_3}$ is consistent with $\nu_{M_3}^{p=1}=3.00(2)$. We have found that $\nu_{s_3}=3.11(3)$ which is also consistent with $\nu_{s_3}^{p=p_c}=\nu_{M_3}^{p=p_c}$~\cite{Lubeck1997BTW}. The full information of the exponents $\tau$ and $\nu$ have been reported in the TABLE \ref{tab:p_c-dist3D} for $M_3, R_3$ and $s_3$. In the first two rows the theoretical and calculated values have been reported for $p=1$ which show a complete agreement. The results of the moment analysis as well as the maximum lattice size have been shown in the last three rows. It is seen that the exponents for $s_3(p=p_c)$ and $M_3(p=p_c)$ are not the same in contrast to the case $p=1$, having its root in the fact that the number of topplings of a typical site in an avalanche is larger than unity for $p=p_c$. Note that for the regular lattice in three and four dimensions these two quantities are nearly the same, showing that the probability that a site topples more than one is very low. Although the resulting exponents differ significantly from the case $p=1$, there are interestingly some similarities with the regular two-dimensional avalanches of the BTW model which has been listed in the third row from the reference \cite{Lubeck1997Numerical}. Note that the difference is seen only for the last one, i.e. $\tau_{s_3}$.\\ 
The fractal dimension $\gamma_{M_3R_3}$ has been shown in Fig. \ref{fig:M3_R3_pc}  for various lattice sizes. We see that this exponent extrapolates to $2.78(8)$ as $L\rightarrow \infty$. This can be interpreted as the effect of empty (un-occupied) sites which lead this exponent to differ from the result for $p=1$, i.e. $\gamma_{M_3R_3}^{p=1}\approx 3$. Observe its similarity to the mass-radius exponent of spanning percolation cluster $\gamma_{M_3R_3}^{\text{percolation}}\approx 2.52$ \cite{isichenko1992percolation}.
\begin{table*}[]
\begin{tabular}{|c|c|c|c|}
\hline $x$ & $M_3$ & $R_3$ & $s_3$  \\
\hline $\tau_x^{p=1,d=3}$(theoretical) & $\frac{4}{3}$ & $2$ & $\approx \tau_{M_3}=\frac{4}{3}$  \\
\hline $\tau_x^{p=1,d=3}$(calculated) & $1.33(3)$ & $1.98(3)$ & $1.34(3)$  \\
\hline $\tau_x^{\text{regular BTW}}(d=2)$ & $1.25$ & $1.59$ & $1.25$  \\
\hline $\nu_{\text{moment anal.}}^{p=p_c,d=3}$ & $3.32(4)$ & $1.12(4)$ & $3.11(3)$ \\
\hline $\tau_{\text{moment anal.}}^{p=p_c,d=3}$ & $1.18(4)$ & $1.49(4)$ & $1.02(3)$ \\
\hline $\tau(L=500)^{p=p_c,d=3}$ & $1.23(3)$ & $1.50(4)$ & $1.05(3)$ \\
\hline
\end{tabular}
\caption{The exponents $\tau$ and $\nu$ of the distribution functions of $M_3,R_3$ and $s_3$ in 3D at the critical regime, i.e. $p=p_c$. The $\tau$ exponents saturate at $\tau_x(L=L_{\text{max}})^{p_c}$ which have reported in the last row. $\tau_x^{p=1,d=3}$(theoretical) and $\tau_x^{p=1,d=3}$(calculated) and $\tau_x^{\text{regular BTW}}(d=2)$ have also been reported for comparison. The results of moment analysis, as well as the maximum lattice size have been shown.} 
\label{tab:p_c-dist3D}
\end{table*}

\begin{table*}[]
\begin{tabular}{|c|c|c|c|c|c|c|}
\hline $x$ & $s_2$ &  $n_{\text{sites}}^{\text{corss-section}}$ & $M_2$ & $r$ & $R_2$ & $a$ \\
\hline $\tau_{\text{moment anal.}}^{p_c}$ & $1.46(9)$ & $1.32(9)$ & $3.00(9)$ & $1.4(8)$ & $3.00(9)$ & $2.39(9)$  \\
\hline $\tau^{\text{Ising}}$ & -- & -- & $2.31$ & -- & $3.4$ & $2.75$  \\
\hline
\end{tabular}
\caption{The exponents of the distribution functions of $s_2, n_{\text{sites}}, M_2, r, R_2$ and $a$ in two-dimensional cross-sections at the critical regime, i.e. $p=p_c$. The exponents have been calculated via the moment analysis. For $\tau^{\text{Ising}}$ the Ref.~\cite{najafi2015observation} has been used.} 
\label{tab:p_c-dist2D}
\end{table*}
\begin{table*}[]
	\begin{tabular}{|c|c|c|c|c|}
		\hline $(x,y)$ & $(M_3,R_3)_{d=3}$ & $(M_2,R_2)_{\text{cross-section}}$ & $(l,r)_{\text{cross-section}}$ & $(l,a)_{\text{cross-section}}$\\
		\hline $\frac{\tau_y^{L\rightarrow\infty}-1}{\tau_x^{L\rightarrow\infty}-1}$ & $2.5(2)$ & $1.1(2)$ & $0.9(1)$ & $0.7(1)$ \\
		\hline $\gamma(L\rightarrow\infty)^{p_c}_{xy}$ & $2.8(1)$ & $1.2(1)$ & $1.37(5)$ & $0.87(3)$ \\
		\hline $\gamma^{\text{2D Ising}}_{xy}$ & -- & -- & $1.375(5)$ & -- \\
		\hline
	\end{tabular}
	\caption{The fractal dimensions $\gamma_{M_3R_3}, \gamma_{M_2R_2}, \gamma_{l,r}$ and $\gamma_{l,a}$ at the critical regime, i.e. $p=p_c$. The observed finite-size relation is $\gamma(L\rightarrow\infty)^{p_c}_{xy}-\gamma(L)^{p_c}_{xy}\propto \frac{\beta_{xy}}{L}$. $\frac{\tau_y^{L\rightarrow\infty}-1}{\tau_x^{L\rightarrow\infty}-1}$ has also been reported for testing the hyper-scaling relation. The same exponents for the Ising model has been shown for comparison, after~\cite{najafi2015observation}.} 
	\label{tab:p_c-FD}
\end{table*}
For the induced model living in the 2D cross-sections (which passes through the center of mass of the spanning cluster), the 2D avalanches are not necessarily connected and for each connected element of 2D avalanche there is an exterior frontier which is a loop $l$ containing the avalanche.\\
The quantities which are analyzed in the cross-sections are the followings:\\
- The mass of 2D avalanches $M2$ which is the total number of sites involved in a 2D cross-section of a avalanche.\\
- The loop lengths $l$ which is the length of the loop that is the external perimeter of a 2D cross-section of a avalanche.\\
- The area inside loops $a$ which is the total area that is contained in the loop which was defined above.\\ 
- The gyration radius of loops $r$ and 2D areas $R_2$ for the cross-sections.\\
- The number of topplings in an avalanche in the cross-section avalanche $s_2$. $s_2$ is the 2D avalanche size, i.e. the total number of topplings in 2D avalanches and $n_{\text{sites}}^{\text{2D}}$ is the number of toppled sites in a connected component of an avalanche.\\
The obtained exponents for two-dimensional cross-sections of avalanches are more interesting which have been reported in TABLE \ref{tab:p_c-dist2D}. The exponents have been obtained, using the moment analysis. The exponent $n_{\text{sites}}$ is compatible with the same exponent of the regular two-dimensional avalanches of the BTW model, i.e. $\tau_{n_{\text{sites}}}(p=p_c) \approx 1.32$. The overall analysis of the exponents in the cross-sections at $p=p_c$ suggest that the model has serious similarities with the 2D critical Ising model. For comparison, the exponents of the Ising model have been shown in the last row of TABLE \ref{tab:p_c-dist2D} for which the similarities are evident. The same is true for the corresponding fractal dimensions. These quantities have been shown in the TABLE \ref{tab:p_c-FD}. $D_F\equiv \gamma_{l,r}$ is interpreted as the most important exponent from which the universality classes of the 2D critical models can be read. The obtained fractal dimension of loops for the 2D cross-section avalanches (2DCSA) is $D_F^{\text{2DCSA}}(p=p_c)=1.37\pm 0.05$ which is compatible with the fractal dimension of the external perimeter of spin clusters of the 2D critical Ising model, i.e. $D_F^{\text{Ising}}=\frac{11}{8}$ \cite{najafi2016monte}. The fact that the fractal dimension $\gamma_{M_2R_2}$ is different substantially from the one for $p=1$ case ( $\gamma_{M_2R_2}^{p=p_c}\approx 1.2$, whereas $\gamma_{M_2R_2}^{p=1}\approx 2$) cannot directly contributed to the existence of these empty sites, since $a$ is the area confined in a loop, i.e. it is purely the effect of the boundaries of the avalanches.\\
The fact that the 2D properties of the model is similar to the Ising model can be roughly understood form the following argument: It is known that the BTW model on the 2D uncorrelated percolation lattice is in the universality class of the Ising model~\cite{najafi2016water,najafi2016bak}. In the other hand, since the percolation system in this study is uncorrelated, the 2D cross sections is also a real sample of 2D percolation lattice. Therefore estimating the cross-sections of the avalanches by the real 2D avalanches, one expects that the critical properties of the BTW model on the 2D percolation lattice is obtained, which is compatible with the Ising universality class. To be more precise about identification of the (3D and 2D) model, we should study its properties out of the percolation threshold, i.e. $p_c<p\leq 1$ which is the subject of the next section.

\section{Out of criticality}\label{off-critical}
In this section we observe how things change in the off-critical regime, i.e. $p_c<p\leq 1$. The quantities to be investigated are the same as the previous section. In the $p_c<p\leq 1$ regime, just like the case $p=p_c$, the critical behaviors are seen with varying exponents to be reported in the next two sub-sections. The results of this part supports the hypothesis that the exponents change logarithmically with respect to $x\equiv p-p_c$ in the off critical regime.
\subsection{Three dimensions}
\begin{figure*}
\begin{subfigure}{0.45\textwidth}\includegraphics[width=\textwidth]{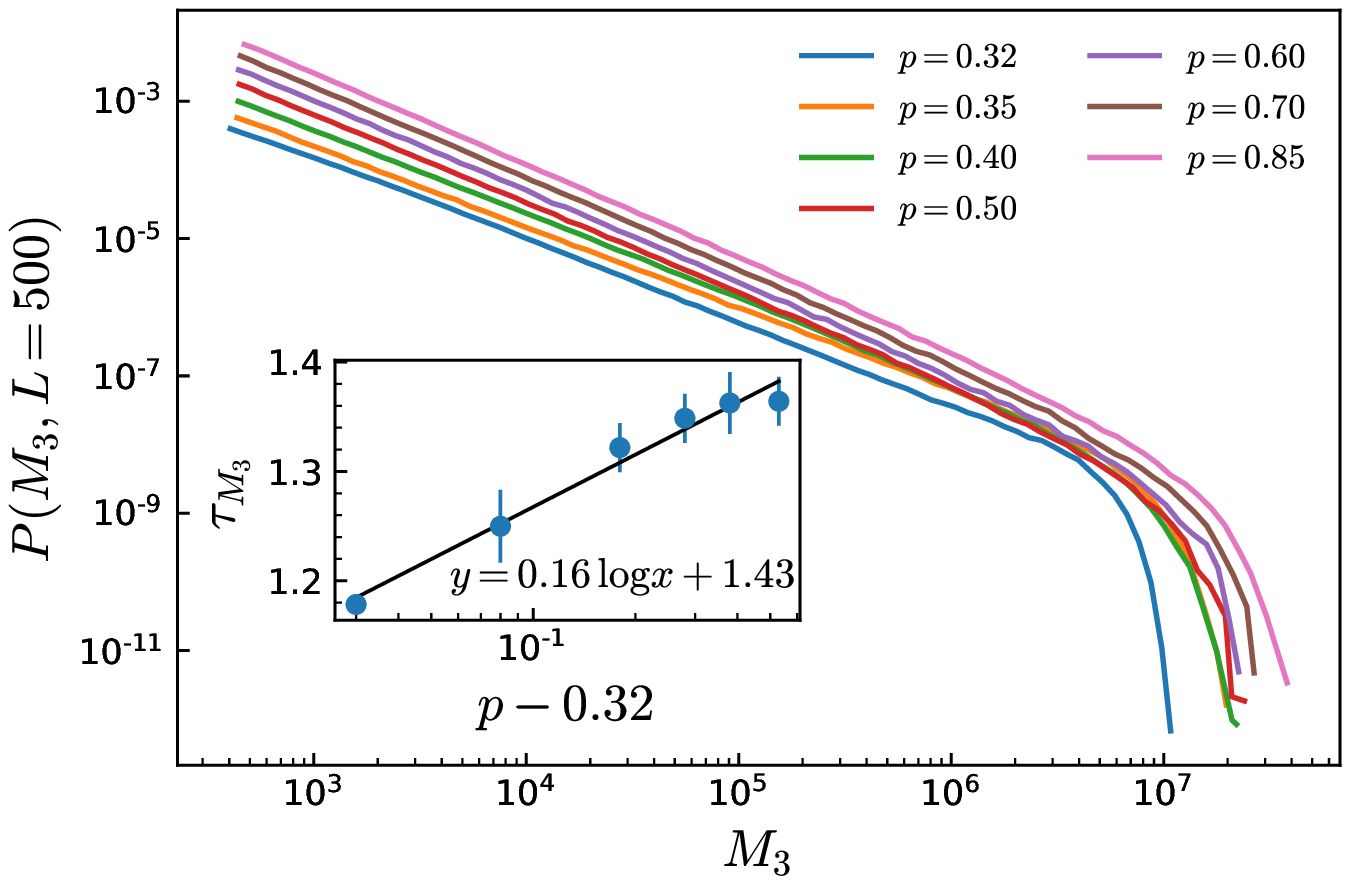}
\caption{}
\label{fig:P_M3_L500}
\end{subfigure}
\begin{subfigure}{0.45\textwidth}\includegraphics[width=\textwidth]{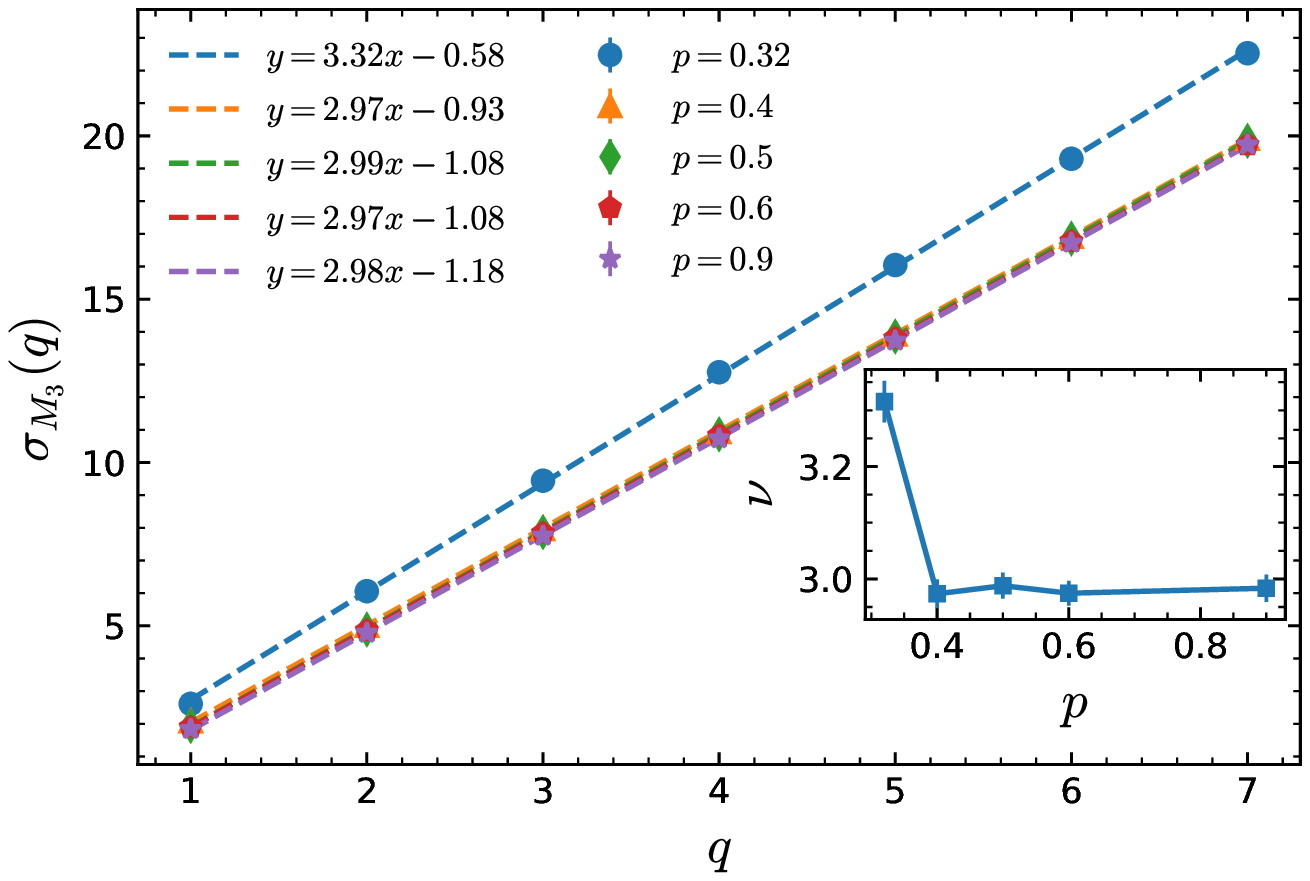}
	\caption{}
	\label{fig:sigma_M3}
\end{subfigure}
\begin{subfigure}{0.45\textwidth}\includegraphics[width=\textwidth]{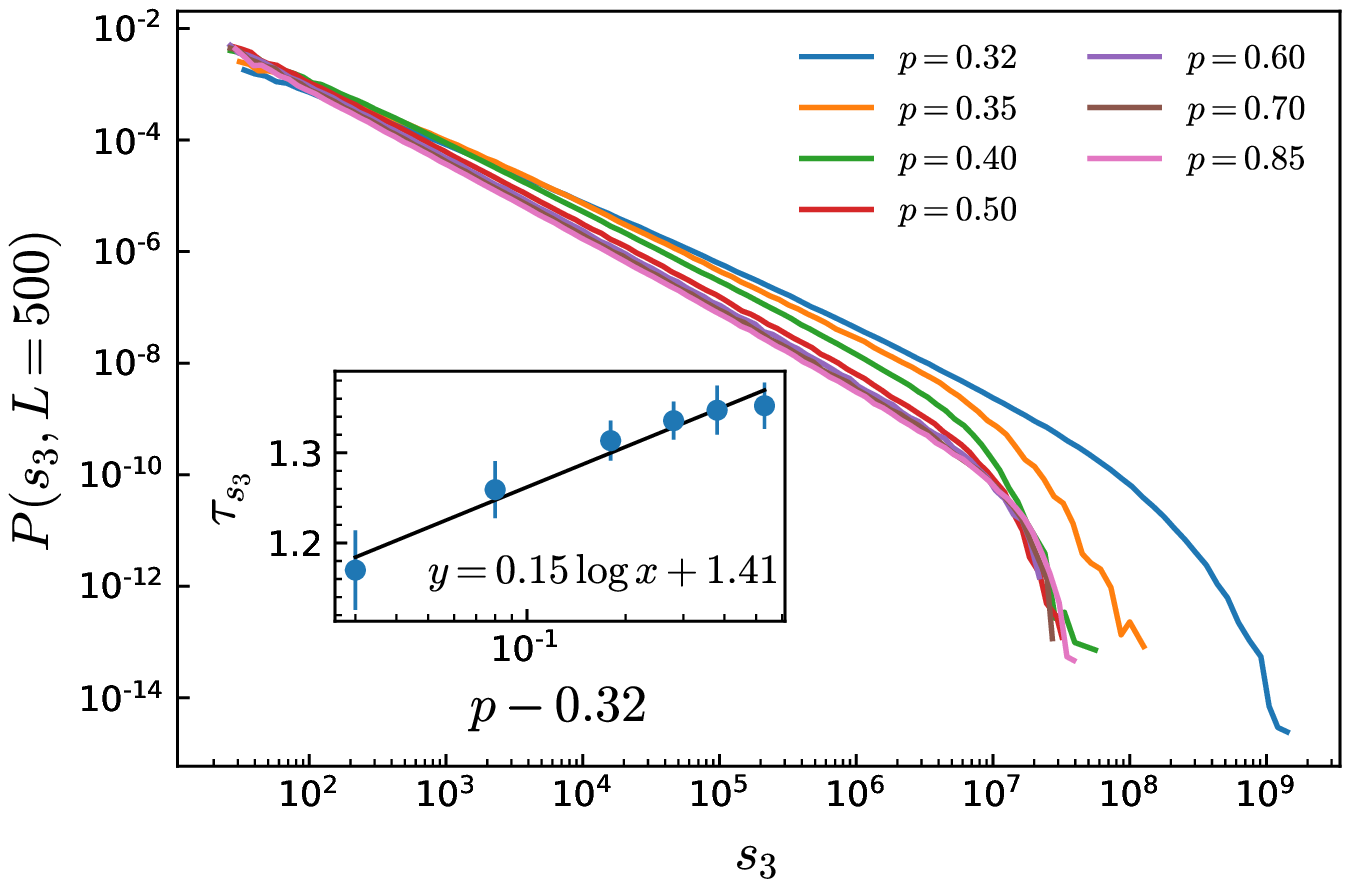}
	\caption{}
	\label{fig:P_s3_L500}
\end{subfigure}
\begin{subfigure}{0.45\textwidth}\includegraphics[width=\textwidth]{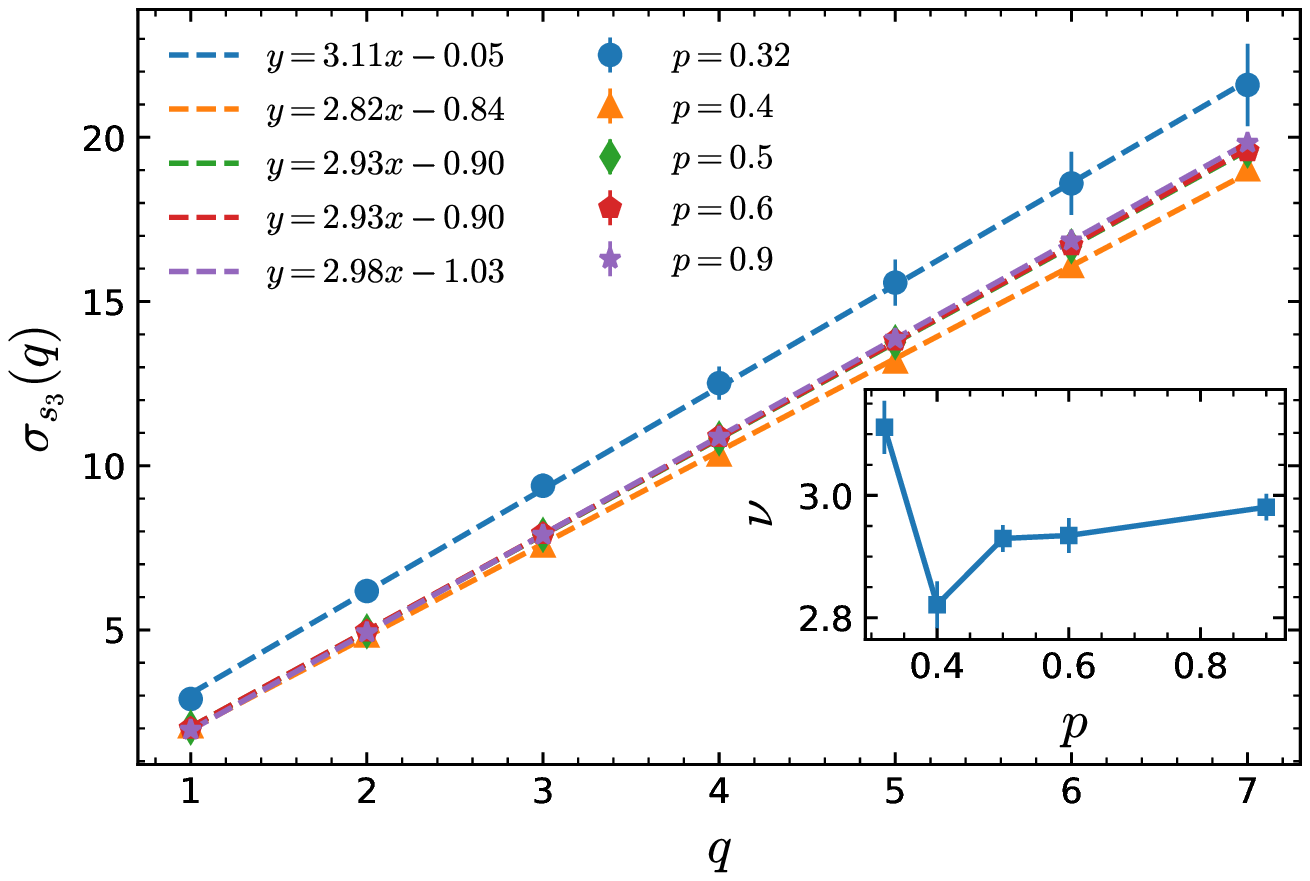}
	\caption{}
	\label{fig:sigma_s3}
\end{subfigure}
\begin{subfigure}{0.45\textwidth}\includegraphics[width=\textwidth]{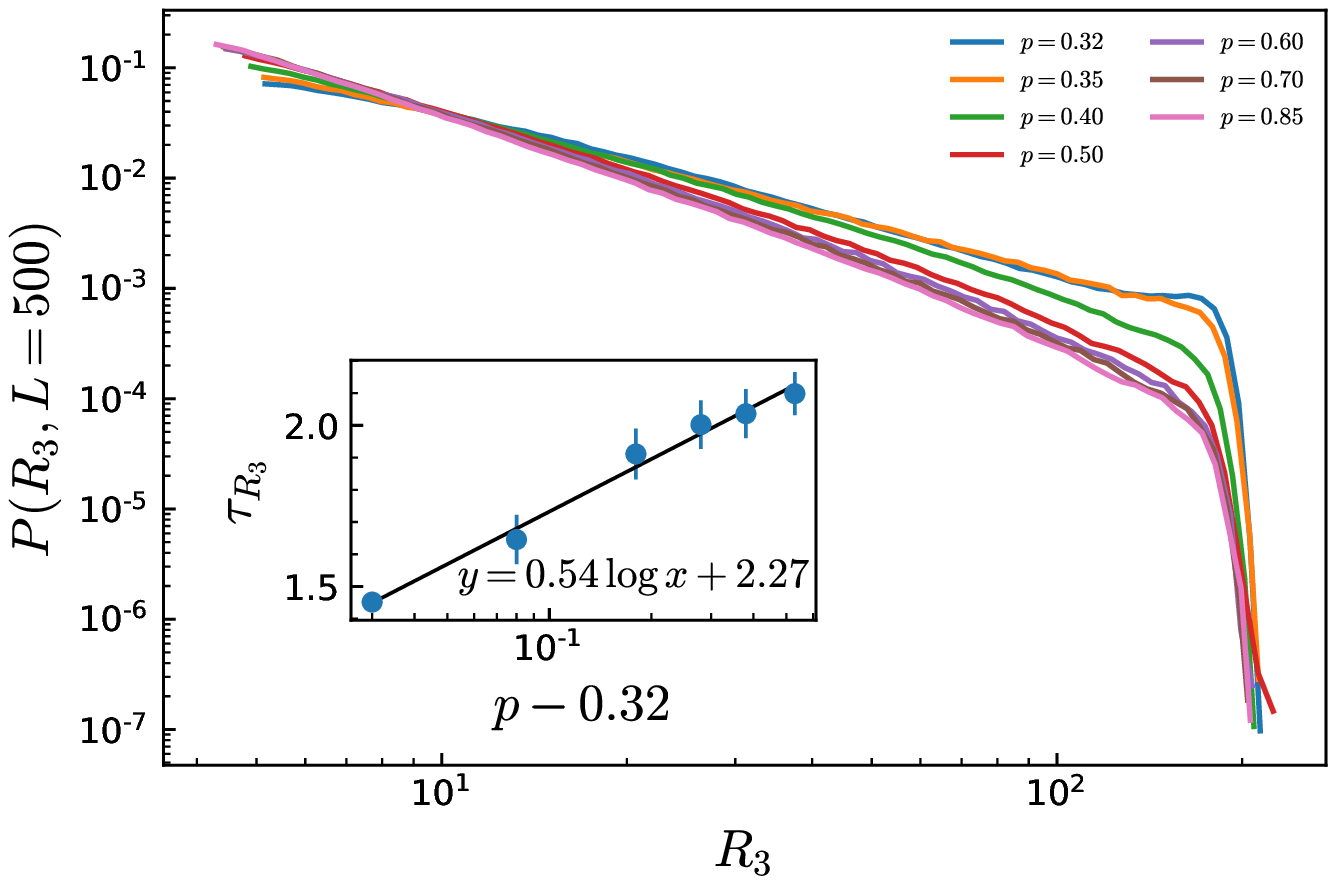}
	\caption{}
	\label{fig:P_R3_L500}
\end{subfigure}
\begin{subfigure}{0.45\textwidth}\includegraphics[width=\textwidth]{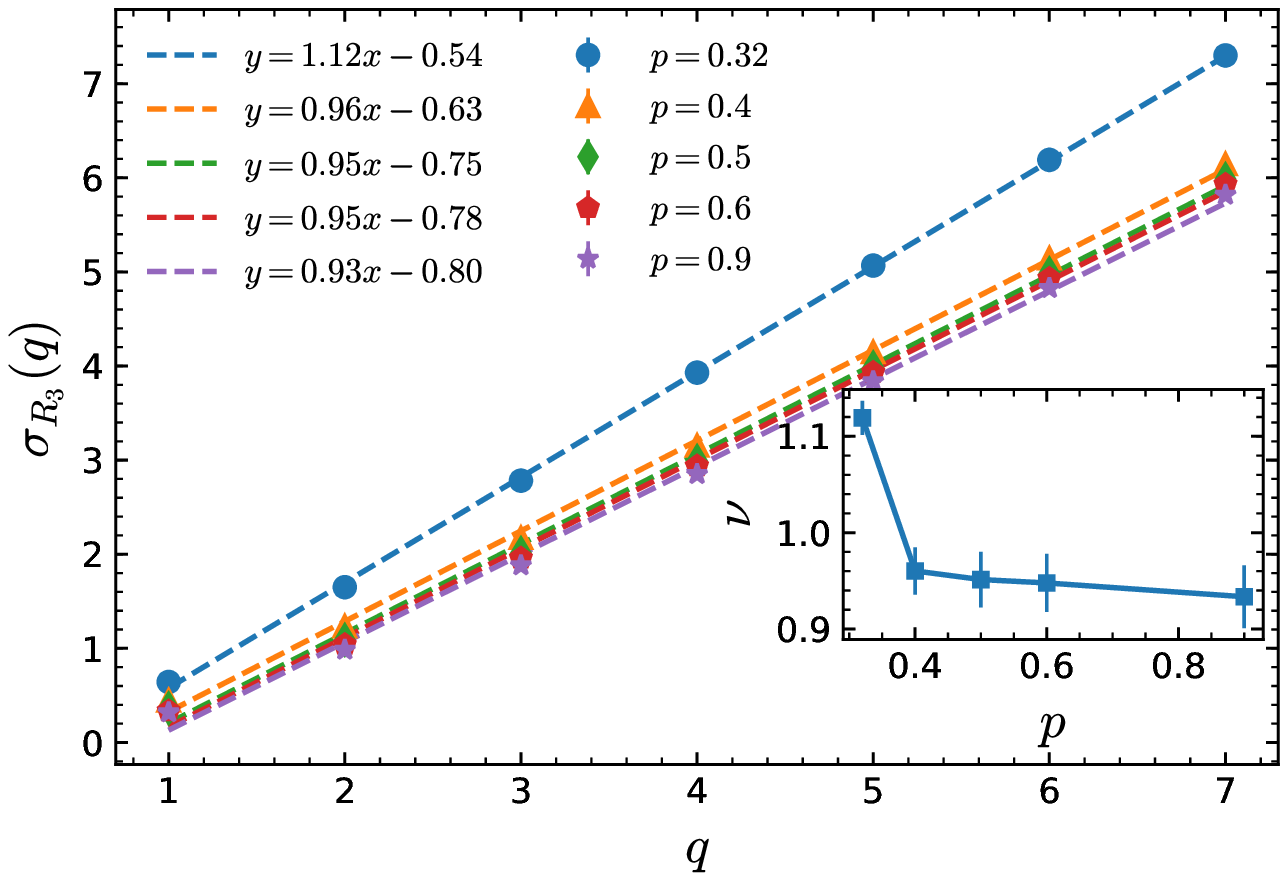}
	\caption{}
	\label{fig:sigma_R3}
\end{subfigure}
\caption{(Color Online) The histogram plot of (a) $M_3$, (c) $s_3$, and (e) $R_3$ in terms of $p$ for $L=500$, along with their moment analysis (b) $\sigma_{M_3}$, (d) $\sigma_{s_3}$, and (f) $\sigma_{R_3}$. The logarithmically-$p$-dependent $\tau$ exponents have been shown in the insets of (a), (c) and (e) graphs ($x\equiv p-p_c$), whereas the $p$-dependent $\nu$ exponents have been shown in the insets of (b), (d) and (f) graphs.}
\label{fig:outcriticality3D}
\end{figure*}
\begin{figure*}
\begin{subfigure}{0.45\textwidth}\includegraphics[width=\textwidth]{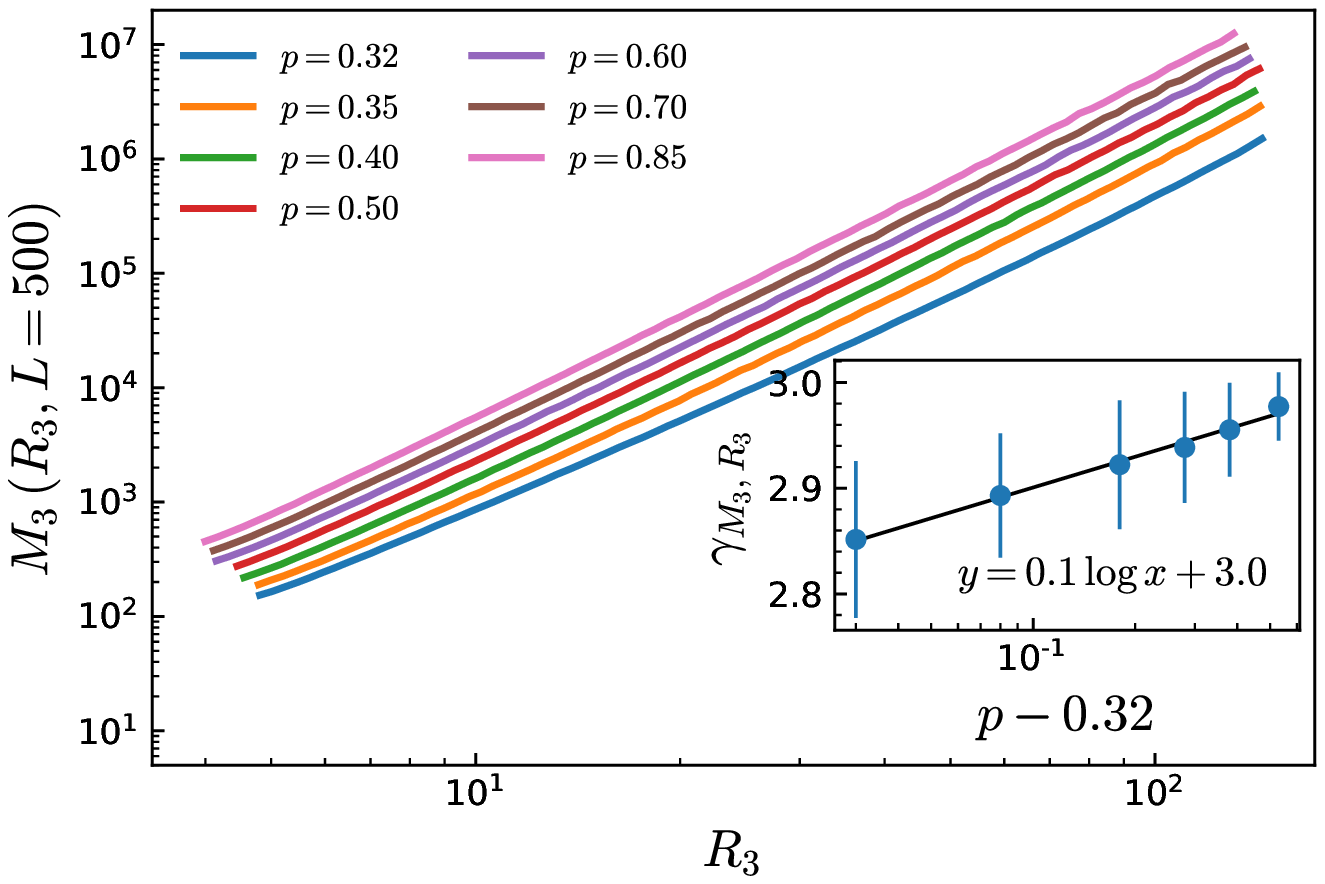}
\caption{}
\label{fig:M3_R3_L500}
\end{subfigure}
\begin{subfigure}{0.45\textwidth}\includegraphics[width=\textwidth]{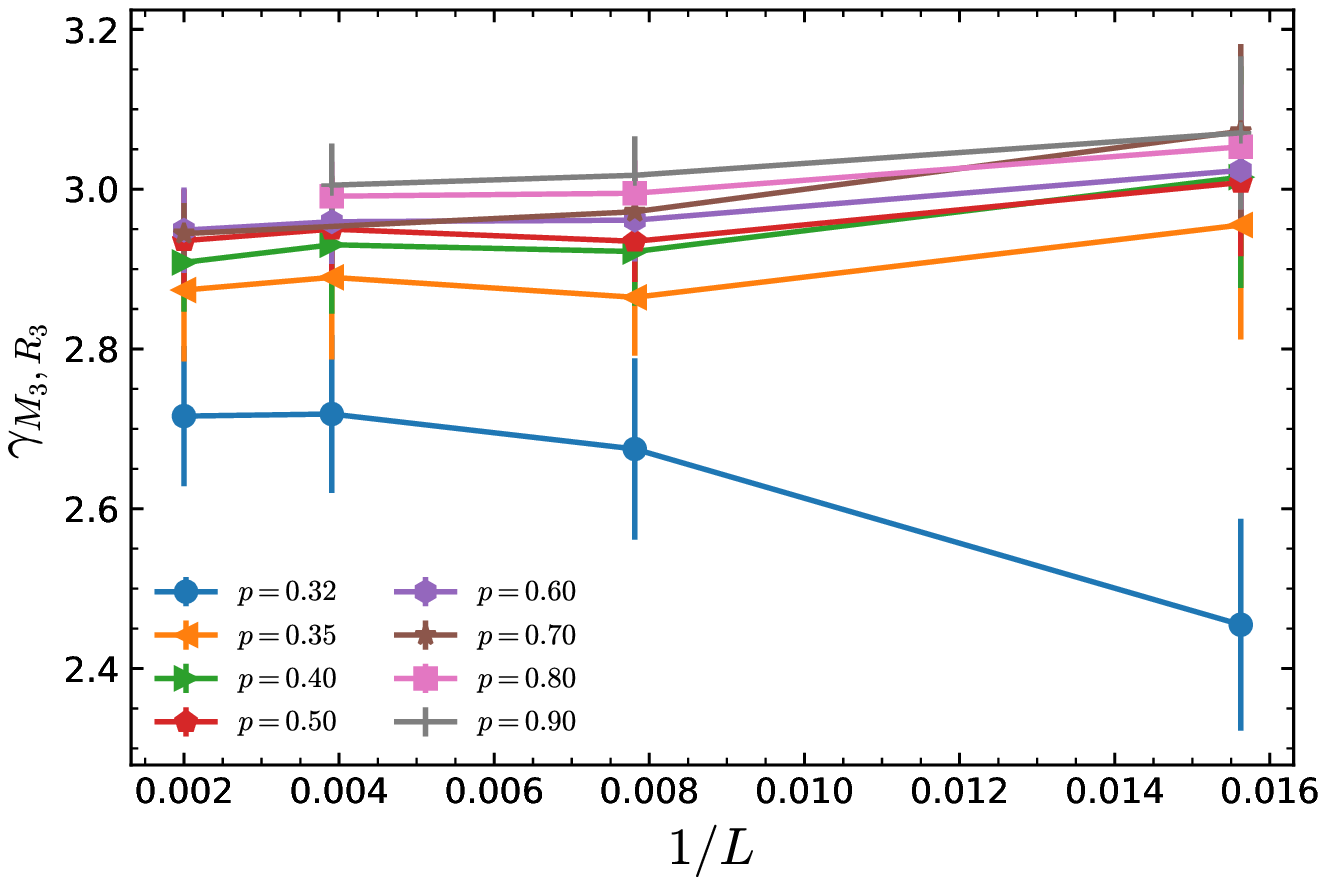}
\caption{}
\label{fig:gamma_M3R3}
\end{subfigure}
\caption{(Color Online) (a) The log-log plot of $M_3$ in terms of $R_3$ with the corresponding $\gamma_{M_3R_3}$ in its inset, which changes logarithmically with $x\equiv p-p_c$. (b) Finite-size dependence of $\gamma_{M_3R_3}$ for various rates of $p$.}
\label{fig:outcriticality3D-FD}
\end{figure*}
For calculating the exponents of the distribution functions of the three-dimensional quantities, along with the direct determining the slopes, we have used the moment analysis for all amounts of $p$. The full information of the graphs have been gathered in Figs~\ref{fig:outcriticality3D} and \ref{fig:outcriticality3D-FD}. In the Fig.\ref{fig:P_M3_L500} we have shown the distribution function of the 3D mass for $L=500$. It is seen that the slopes change smoothly from $\tau_{M_3}(p_c)$ to $\tau_{M_3}(p=1)$. By the moment analysis, in addition to extracting the exponents, we can calculate the $\nu$ exponent which controls the cut value (e.g. $R_3^{\text{cut}}$ for $R_3$). In the Fig.~\ref{fig:sigma_M3} we have shown $\sigma_{M_3}(q)$ and $\nu_{M_3}$ in its inset. Our results reveal that all of the $\nu$ exponents fall off rapidly from its value in $p=p_c$ to that of $p>p_c$ which is nearly constant (see Figs.~\ref{fig:sigma_M3}, \ref{fig:sigma_s3} and \ref{fig:sigma_R3}). \\
As mentioned above, the important feature of the results for 3D is that no cross-over between two regimes, i.e. \textit{UV} (small-scale) regime and \textit{IR} (large-scale) regime, is seen. Instead $\tau_x$ varies linearly with the logarithm of $p-p_c$, i.e. $\tau_x(p)=\zeta_x\ln(p-p_c)+\xi_x$ (note that apparently this relation is not valid for $p$ very close to $p_c$). The same is true for $s_3$ in the Fig. \ref{fig:P_s3_L500}. This behavior is seen for all lattice sizes considered in this paper with $L$-dependent $\zeta_x$ and $\xi_x$. The resulting coefficients although do not have a clean scaling behavior in terms of $1/L$ or $1/\ln L$, but saturate properly for $L\gtrsim 256$. Therefore in the TABLE \ref{tab:3d-tau} we have reported the results for $L=500$. The same behavior is seen for $\gamma_{M_3R_3}$ for which a logarithmic behavior in terms of $p-p_c$ is seen (the inset of Fig. \ref{fig:M3_R3_L500}). The finite size dependence of the $\gamma_{M_3R_3}$ for various occupation numbers ($p$) has been shown in Fig. \ref{fig:gamma_M3R3} from which we see that the exponents become nearly saturated for large sizes.  It is worth mentioning that the hyper-scaling relation $\gamma_{M_3R_3}\equiv \frac{\tau_{R_3}-1}{\tau_{M_3}-1}$ is violated for $p_c<p<1$ and is restored right at $p=p_c$ and $p=1$.

\begin{table}[]
\begin{tabular}{|c|c|c|c|}
\hline $x$ & $M_3$ &  $s_3$ &  $R_3$  \\
\hline $\zeta_x$ & $0.16(4)$ & $0.15(4)$ & $0.54(5)$ \\
\hline $\xi_x$ & $1.43(5)$ & $1.41(5)$ & $2.27(5)$ \\
\hline
\end{tabular}
\caption{The coefficients of the relation $\tau_x(p)=\zeta_x\ln(p-p_c)+\xi_x$ for the exponents of the distribution functions in three dimensions.} 
\label{tab:3d-tau}
\end{table}

\subsection{Two dimensions; cross-section statistics}
\begin{figure*}
\begin{subfigure}{0.45\textwidth}\includegraphics[width=\textwidth]{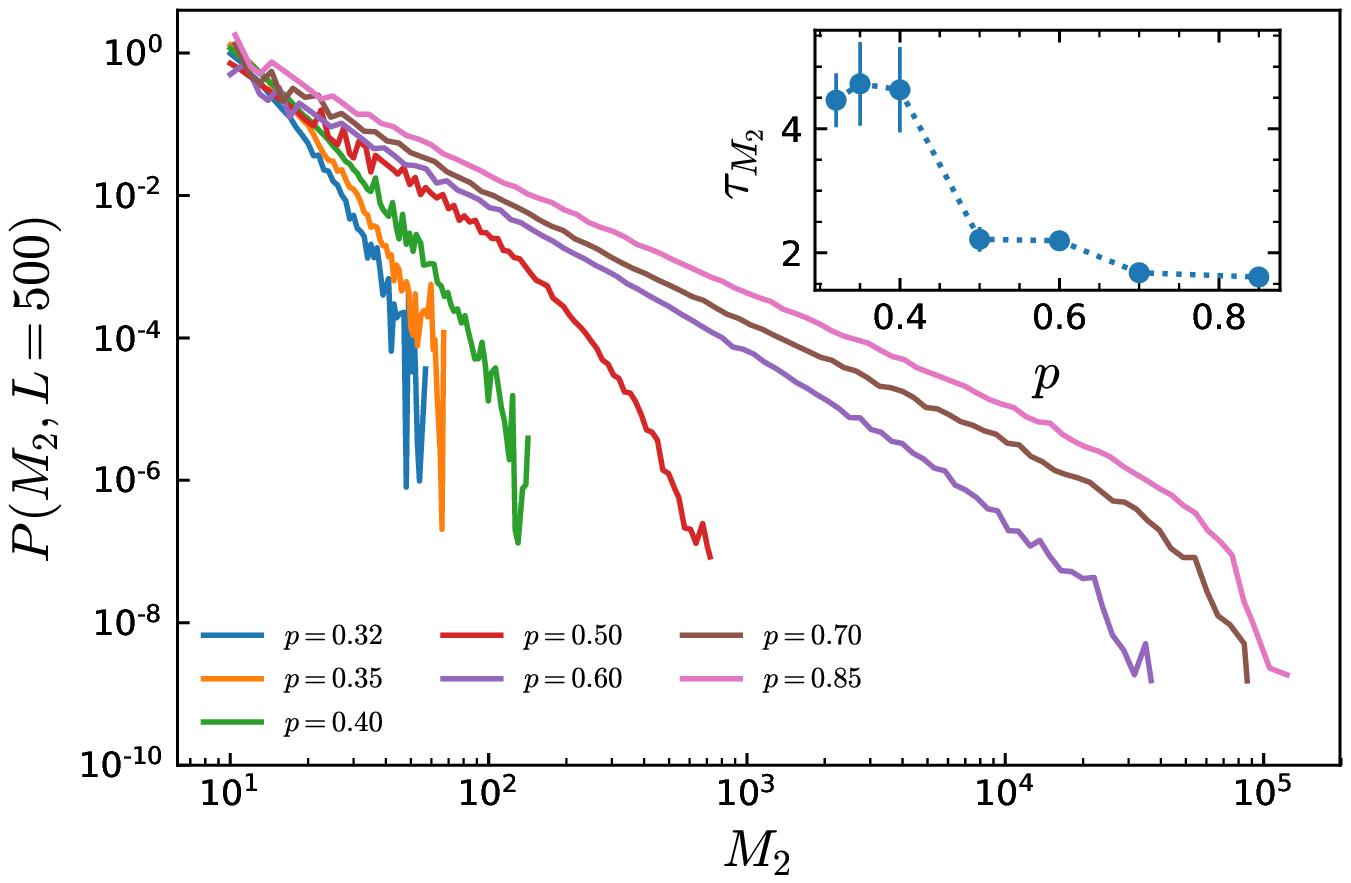}
\caption{}
\label{fig:P_M2_L500}
\end{subfigure}
\begin{subfigure}{0.45\textwidth}\includegraphics[width=\textwidth]{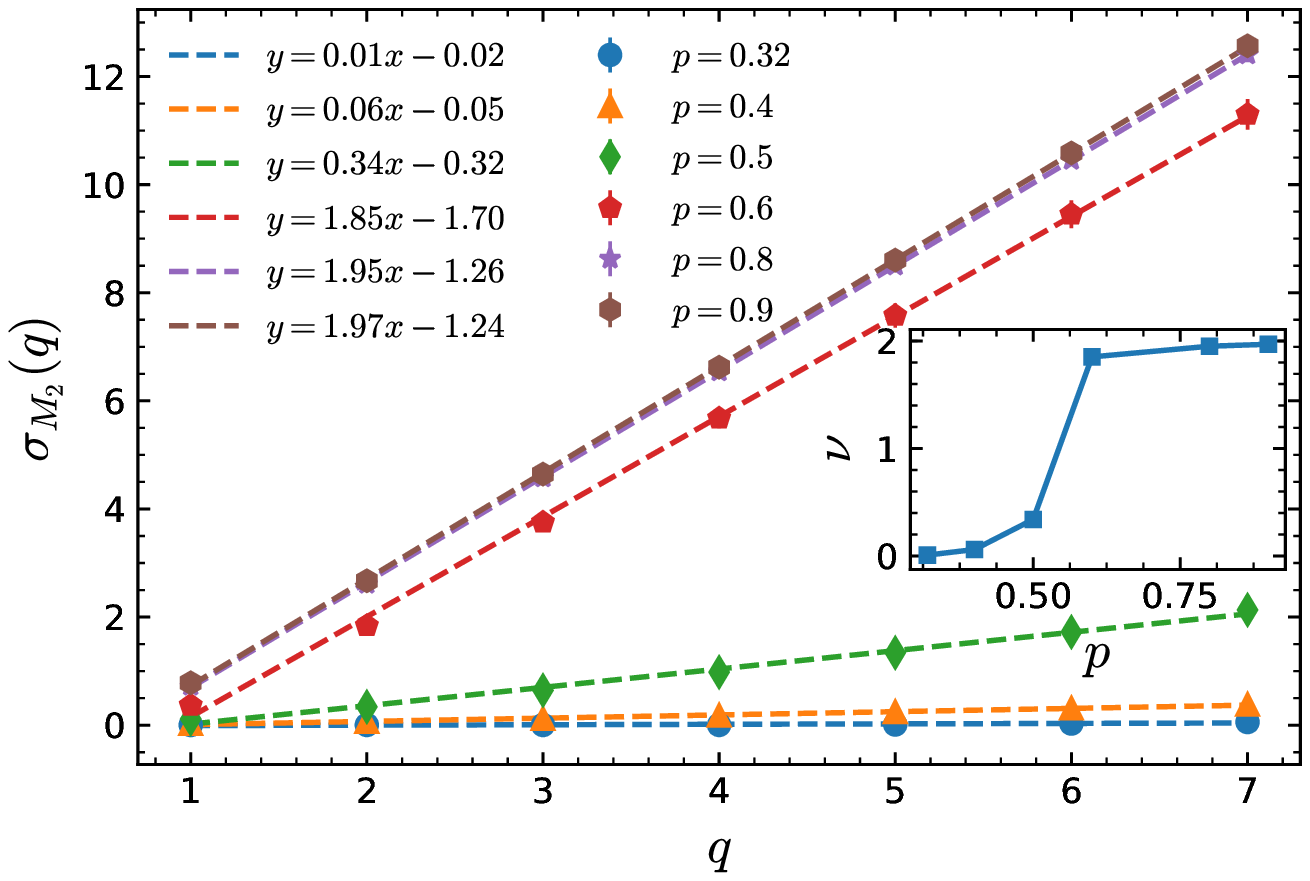}
	\caption{}
	\label{fig:sigma_M2}
\end{subfigure}
\begin{subfigure}{0.45\textwidth}\includegraphics[width=\textwidth]{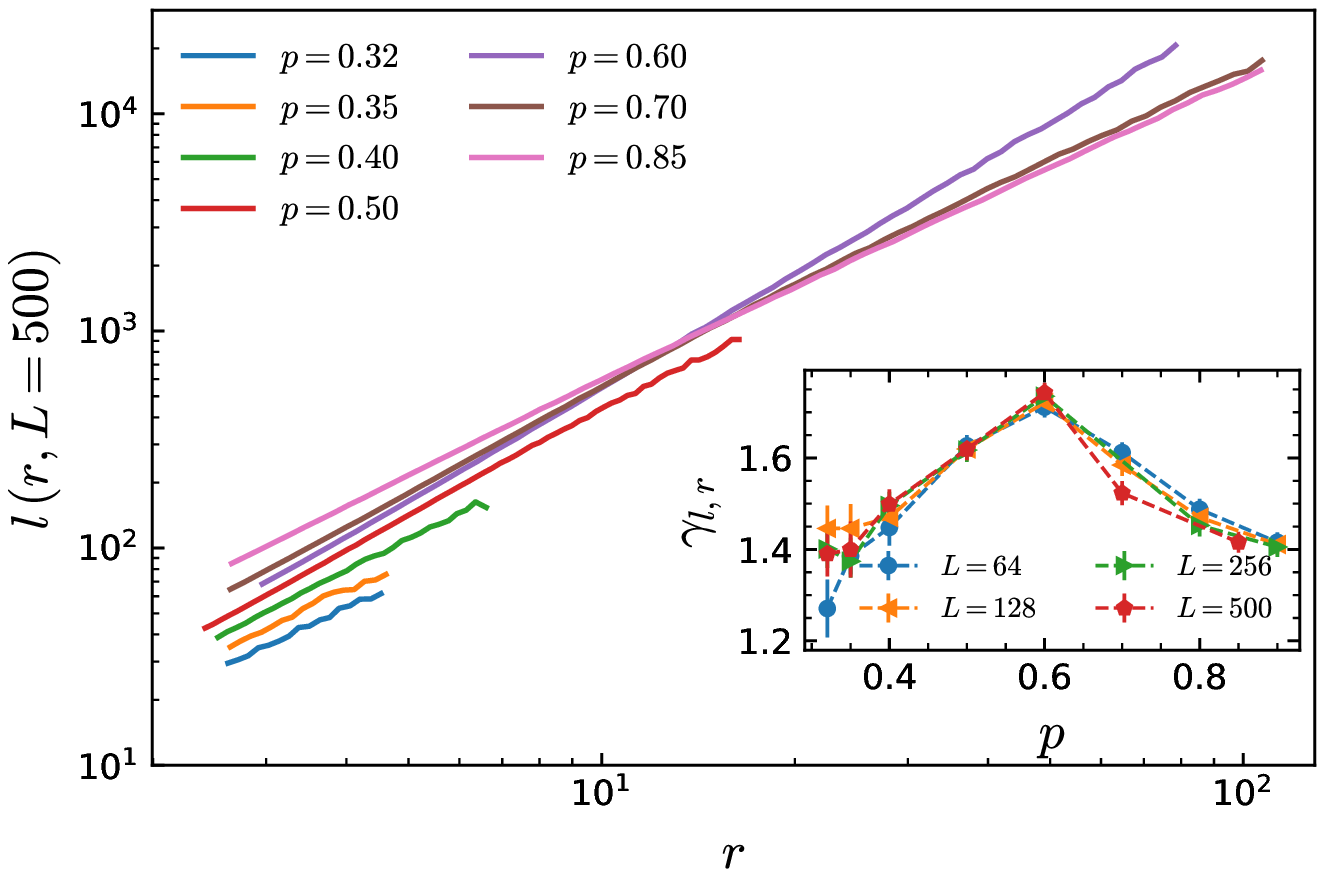}
	\caption{}
	\label{fig:lr_L500}
\end{subfigure}
\begin{subfigure}{0.45\textwidth}\includegraphics[width=\textwidth]{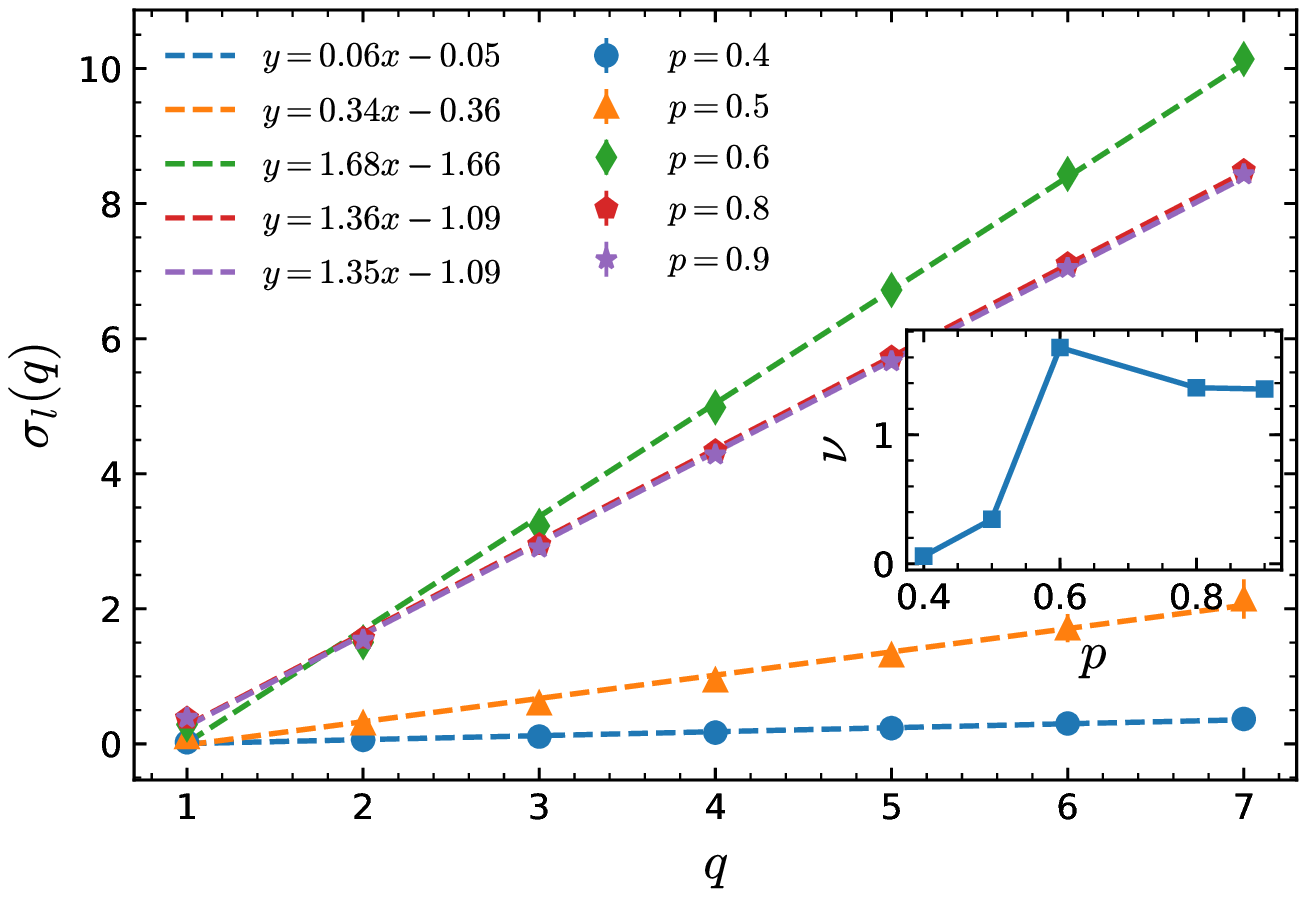}
	\caption{}
	\label{fig:sigma_l}
\end{subfigure}
\begin{subfigure}{0.45\textwidth}\includegraphics[width=\textwidth]{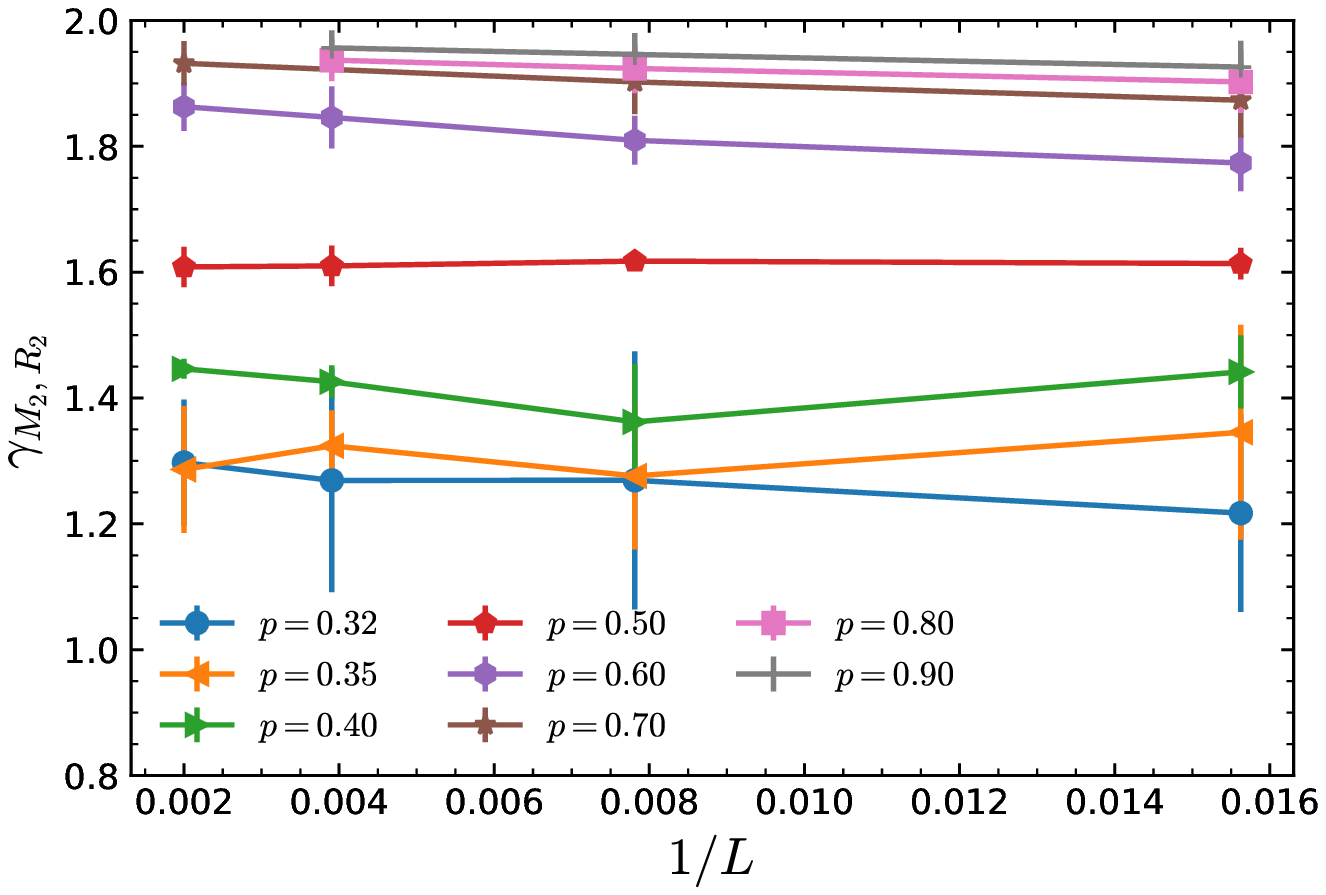}
	\caption{}
	\label{fig:gamma_M2R2}
\end{subfigure}
\begin{subfigure}{0.45\textwidth}\includegraphics[width=\textwidth]{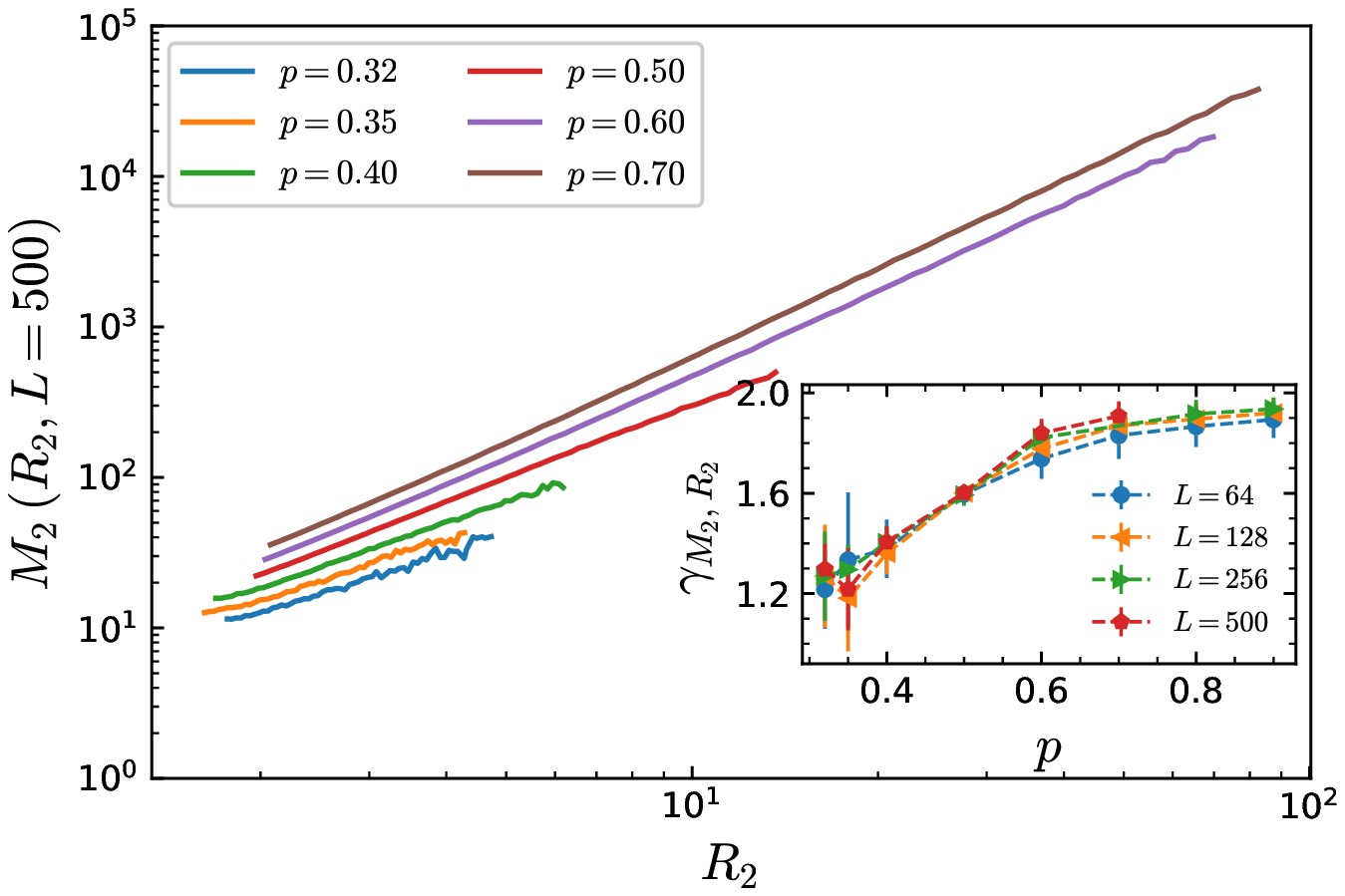}
	\caption{}
	\label{fig:M2R2_L500}
\end{subfigure}
\caption{(Color Online) (a) The histogram plot of $M_2$ (for cross-sections) in terms of $p$ with their exponents (insets) for $L=500$. (b) $\sigma_{M_2}$ in terms of $p$ with its corresponding $\nu$ exponent in the inset. (c) The fractal dimension $\gamma_{l,r}$ in terms of $p$. Inset shows the behavior of $\gamma_{l,r}$ in terms of $p$ which shows a maximum at $p=p_0\approx 0.6$. (d) $\sigma_{l}$ in terms of $p$ with its corresponding $\nu$ exponent in the inset. (e) Finite-size dependence of $\gamma_{M_2R_2}$ for various rates of $p$. (f) The log-log plot of $M_2$ in terms of $R_2$ for various amounts of $p$ for $L=500$. Inset shows the $p$ dependence of $\gamma_{M_2R_2}$.}
\label{fig:outcriticality2D}
\end{figure*}

Now let us turn to the two-dimensional problem in the $p_c<p\leq 1$ regime. The two dimensional problem is interesting since a singular behavior is seen in some occupation number. The BTW model on the two-dimensional site-diluted square lattice has been studied in some previous works \cite{najafi2016bak}. This problem has proved to have some relations with more empirical models \cite{najafi2016water}. A very interesting feature of this study (Ref. \cite{najafi2016water}) is highlighting the point that there is a special amount of occupation number $p_0$ for which the percolation probability is maximum. This quantity is in the vicinity of $p=0.6$. Interestingly we see something like this in the 2D induced model, i.e. the model living in the 2D cross-sections. For most considered quantities, we observed a singular behavior around $p_0\in (0.5,0.6)$. To show this effect we have sketched the log-log plot of the distribution function of the 2D avalanche mass $P(M_2)$ in the Fig. \ref{fig:P_M2_L500} for various rates of $p$. In its inset we have shown the quantity $\tau_{M_2}$ in terms of $p$. For small value of $p-p_c$ the exponent is nearly constant up to $p=0.4$ at which the exponent falls rapidly to another constant which finally saturates at $p=1$. This behavior occurs also for most of the two-dimensional observables considered in this work. Consider for example the $\nu$ exponent in the inset of Fig.~\ref{fig:sigma_M2} in which a cross-over is seen from small to large values in terms of $p$. As another example let us consider the fractal dimension $\gamma_{lr}$ which has been presented in Fig. \ref{fig:lr_L500}, whose inset represents the exponent for various lattice sizes. The mentioned singular behavior is seen in a sharp peak at $p=0.6$. An explanation is the fact that the 2D cross section of a 3D uncorrelated percolation lattice is a real 2D percolation lattice. For the 2D square percolation lattice we know that $p_c^{2D}=0.5927$ in which a percolation transition occurs. So for $p_c\leq p\leq p_c^{2D}$ we have no \textit{two-dimensional} percolated cluster, i.e. when we are restricted to the cross sections, this phase is non-percolating one. Therefore the 2D behaviors are separated into two distinct phases: $p_c\leq p\leq p_c^{2D}$ and $p_c^{2D}<p\leq 1$. For the former case the 2D clusters are surely non-percolating and for the latter case the clusters have the chance to be percolating. This is responsible for the observed singularity at $p=p_c^{2D}$. This behavior can also be seen in the inset figure \ref{fig:sigma_l}, in which $\nu_l$ crosses over from small $p$'s to the large ones, and also the inset of Fig.~\ref{fig:M2R2_L500} in which $\gamma_{M_2R_2}$ do such a cross over. \\
Based on the results, one may claim that there are two separate phases at least for the model on the cross-sections, i.e. $(p_c,p_0)$ and $(p_0,1)$ ($p_0\simeq p_c^{2D}$) each of which has its own off-critical behaviors. Although this effect is seen in finite size samples, we see from the inset of \ref{fig:lr_L500} that the change of slope is not $L$-dependent for the sample sizes considered in this work. More precise results can be obtained for larger sample sizes. The fact that $p_0$ is $L$-dependent or not cannot be deduced from the lattice sizes considered in this work. \\
We said that in the thermodynamic limit all of the properties of the system for $p_c<p\leq 1$ become identical to the $p=1$ case. To be more precise, we should mention that the phase $p_c^{3D}\leq p<p_c^{2D}$ in the cross sections has no thermodynamic limit, since the 2D samples have not percolated in this limit. Note that by this 2D cluster we mean the cluster which is restricted to cross sections, and should not confused with the original 3D cluster is which surely percolated. We expect however that all $p\geq p_c^{2D}$ tend to $p=1$ statistics in the thermodynamic limit in the cross-sections, since the clusters have the chance to percolate over 2D system in this limit. This is properly seen in Fig.~\ref{fig:M2R2_L500} from which it is seen that the $\gamma_{M_2R_2}$ exponent runs by increasing $L$ for $p\geq p_c^{2D}$ in such a way that the portion of the $\gamma_{M_2R_2}$-graph which is close to the $\gamma_{M_2R_2}(p=1)$ become more and more wide, i.e. the point at which the nearly linear increase of $\gamma_{M_2R_2}$ is changed becomes smaller for higher lattice sizes. This shows that the $p=1$ properties of the system becomes dominant for $L\rightarrow\infty$ which is expected. The finite size dependence of this exponent has been shown in Fig. \ref{fig:gamma_M2R2}, which reveals the approaching of the exponents to the thermodynamic limit. The figure shows that all exponents are nearly saturated for $L\gtrsim 256$ as stated above.\\
We conclude that although the $p$-dependence of the exponents in 3D is logarithmic, in 2D cross-sections the dependence is not as simple and has some features, e.g. there is a singular behavior at $p=p_0$.

\section{Conclusion}
\label{conclusion}

In this paper we have considered the three-dimensional BTW model on the uncorrelated site-diluted cubic percolation lattice which is tuned by the occupation number $p$. 
Along with the original lattice, we have also considered the two-dimensional cross-sections of the system which crosses the center of mass of the spanning cluster. Our motivation for this has been to investigate the energy propagation in lower dimensional ($d-1=2$) system affected by the original ($d=3$) lattice, which we name the 2D induced model. This is of both theoretical and empirical interest. We had two separate studies: critical $p=p_c$ and off-critical $p_c<p\leq 1$ regimes and the fractal dimensions and the distribution functions of various statistical observables have been studied vie the moment analysis. For the critical case some proper finite-size scaling were observed and some resulting exponents were observed to be compatible with 2D BTW model. The exponents of the quantities in 2D cross-sections are compatible with the 2D Ising universality class. These exponents satisfy also some hyper-scaling relations. For the off-critical case in three dimensions we have observed that the exponents change logarithmically with $p-p_c$ violating the hyper-scaling relations obtained for the critical case. For the 2D induced model in the off-critical regime we showed that there is a $p$ value ($p_0\in (0.5,0.6)$) at which the behavior of the system changes. This is reminiscent of the previously observed occupation number at which the percolation probability becomes maximum in the BTW model on the 2D site-diluted percolation lattice \cite{najafi2016bak}. We conclude that the system for $p_c^{3D}\leq p<p_c^{2D}$ in the cross-sections does not have a thermodynamic limit, whereas for $p\geq p_c^{2D}$ the system is identical to the $p=1$ system.

\bibliography{refs}

\begin{thebibliography}{50}%
\makeatletter
\providecommand \@ifxundefined [1]{%
 \@ifx{#1\undefined}
}%
\providecommand \@ifnum [1]{%
 \ifnum #1\expandafter \@firstoftwo
 \else \expandafter \@secondoftwo
 \fi
}%
\providecommand \@ifx [1]{%
 \ifx #1\expandafter \@firstoftwo
 \else \expandafter \@secondoftwo
 \fi
}%
\providecommand \natexlab [1]{#1}%
\providecommand \enquote  [1]{``#1''}%
\providecommand \bibnamefont  [1]{#1}%
\providecommand \bibfnamefont [1]{#1}%
\providecommand \citenamefont [1]{#1}%
\providecommand \href@noop [0]{\@secondoftwo}%
\providecommand \href [0]{\begingroup \@sanitize@url \@href}%
\providecommand \@href[1]{\@@startlink{#1}\@@href}%
\providecommand \@@href[1]{\endgroup#1\@@endlink}%
\providecommand \@sanitize@url [0]{\catcode `\\12\catcode `\$12\catcode
  `\&12\catcode `\#12\catcode `\^12\catcode `\_12\catcode `\%12\relax}%
\providecommand \@@startlink[1]{}%
\providecommand \@@endlink[0]{}%
\providecommand \url  [0]{\begingroup\@sanitize@url \@url }%
\providecommand \@url [1]{\endgroup\@href {#1}{\urlprefix }}%
\providecommand \urlprefix  [0]{URL }%
\providecommand \Eprint [0]{\href }%
\providecommand \doibase [0]{http://dx.doi.org/}%
\providecommand \selectlanguage [0]{\@gobble}%
\providecommand \bibinfo  [0]{\@secondoftwo}%
\providecommand \bibfield  [0]{\@secondoftwo}%
\providecommand \translation [1]{[#1]}%
\providecommand \BibitemOpen [0]{}%
\providecommand \bibitemStop [0]{}%
\providecommand \bibitemNoStop [0]{.\EOS\space}%
\providecommand \EOS [0]{\spacefactor3000\relax}%
\providecommand \BibitemShut  [1]{\csname bibitem#1\endcsname}%
\let\auto@bib@innerbib\@empty
%</preamble>
\bibitem [{\citenamefont {Kose}\ \emph {et~al.}(2009)\citenamefont {Kose},
  \citenamefont {Fischer}, \citenamefont {Mao},\ and\ \citenamefont
  {Koser}}]{kose2009label}%
  \BibitemOpen
  \bibfield  {author} {\bibinfo {author} {\bibfnamefont {A.~R.}\ \bibnamefont
  {Kose}}, \bibinfo {author} {\bibfnamefont {B.}~\bibnamefont {Fischer}},
  \bibinfo {author} {\bibfnamefont {L.}~\bibnamefont {Mao}}, \ and\ \bibinfo
  {author} {\bibfnamefont {H.}~\bibnamefont {Koser}},\ }\href@noop {}
  {\bibfield  {journal} {\bibinfo  {journal} {Proceedings of the National
  Academy of Sciences}\ }\textbf {\bibinfo {volume} {106}},\ \bibinfo {pages}
  {21478} (\bibinfo {year} {2009})}\BibitemShut {NoStop}%
\bibitem [{\citenamefont {Kikura}\ \emph {et~al.}(2004)\citenamefont {Kikura},
  \citenamefont {Matsushita}, \citenamefont {Matsuzaki}, \citenamefont
  {Kobayashi},\ and\ \citenamefont {Aritomi}}]{kikura2004thermal}%
  \BibitemOpen
  \bibfield  {author} {\bibinfo {author} {\bibfnamefont {H.}~\bibnamefont
  {Kikura}}, \bibinfo {author} {\bibfnamefont {J.}~\bibnamefont {Matsushita}},
  \bibinfo {author} {\bibfnamefont {M.}~\bibnamefont {Matsuzaki}}, \bibinfo
  {author} {\bibfnamefont {Y.}~\bibnamefont {Kobayashi}}, \ and\ \bibinfo
  {author} {\bibfnamefont {M.}~\bibnamefont {Aritomi}},\ }\href@noop {}
  {\bibfield  {journal} {\bibinfo  {journal} {Science and Technology of
  Advanced Materials}\ }\textbf {\bibinfo {volume} {5}},\ \bibinfo {pages}
  {703} (\bibinfo {year} {2004})}\BibitemShut {NoStop}%
\bibitem [{\citenamefont {Matsuzaki}\ \emph {et~al.}(2004)\citenamefont
  {Matsuzaki}, \citenamefont {Kikura}, \citenamefont {Matsushita},
  \citenamefont {Aritomi},\ and\ \citenamefont {Akatsuka}}]{matsuzaki2004real}%
  \BibitemOpen
  \bibfield  {author} {\bibinfo {author} {\bibfnamefont {M.}~\bibnamefont
  {Matsuzaki}}, \bibinfo {author} {\bibfnamefont {H.}~\bibnamefont {Kikura}},
  \bibinfo {author} {\bibfnamefont {J.}~\bibnamefont {Matsushita}}, \bibinfo
  {author} {\bibfnamefont {M.}~\bibnamefont {Aritomi}}, \ and\ \bibinfo
  {author} {\bibfnamefont {H.}~\bibnamefont {Akatsuka}},\ }\href@noop {}
  {\bibfield  {journal} {\bibinfo  {journal} {Science and Technology of
  Advanced Materials}\ }\textbf {\bibinfo {volume} {5}},\ \bibinfo {pages}
  {667} (\bibinfo {year} {2004})}\BibitemShut {NoStop}%
\bibitem [{\citenamefont {Philip}\ \emph {et~al.}(2007)\citenamefont {Philip},
  \citenamefont {Shima},\ and\ \citenamefont {Raj}}]{philip2007enhancement}%
  \BibitemOpen
  \bibfield  {author} {\bibinfo {author} {\bibfnamefont {J.}~\bibnamefont
  {Philip}}, \bibinfo {author} {\bibfnamefont {P.}~\bibnamefont {Shima}}, \
  and\ \bibinfo {author} {\bibfnamefont {B.}~\bibnamefont {Raj}},\ }\href@noop
  {} {\bibfield  {journal} {\bibinfo  {journal} {Applied physics letters}\
  }\textbf {\bibinfo {volume} {91}},\ \bibinfo {pages} {203108} (\bibinfo
  {year} {2007})}\BibitemShut {NoStop}%
\bibitem [{\citenamefont {Kim}\ \emph {et~al.}(2008)\citenamefont {Kim},
  \citenamefont {Fang}, \citenamefont {Choi},\ and\ \citenamefont
  {Seo}}]{kim2008magnetic}%
  \BibitemOpen
  \bibfield  {author} {\bibinfo {author} {\bibfnamefont {J.~H.}\ \bibnamefont
  {Kim}}, \bibinfo {author} {\bibfnamefont {F.~F.}\ \bibnamefont {Fang}},
  \bibinfo {author} {\bibfnamefont {H.~J.}\ \bibnamefont {Choi}}, \ and\
  \bibinfo {author} {\bibfnamefont {Y.}~\bibnamefont {Seo}},\ }\href@noop {}
  {\bibfield  {journal} {\bibinfo  {journal} {Materials Letters}\ }\textbf
  {\bibinfo {volume} {62}},\ \bibinfo {pages} {2897} (\bibinfo {year}
  {2008})}\BibitemShut {NoStop}%
\bibitem [{\citenamefont {Benkoski}\ \emph {et~al.}(2008)\citenamefont
  {Benkoski}, \citenamefont {Bowles}, \citenamefont {Jones}, \citenamefont
  {Douglas}, \citenamefont {Pyun},\ and\ \citenamefont
  {Karim}}]{benkoski2008self}%
  \BibitemOpen
  \bibfield  {author} {\bibinfo {author} {\bibfnamefont {J.~J.}\ \bibnamefont
  {Benkoski}}, \bibinfo {author} {\bibfnamefont {S.~E.}\ \bibnamefont
  {Bowles}}, \bibinfo {author} {\bibfnamefont {R.~L.}\ \bibnamefont {Jones}},
  \bibinfo {author} {\bibfnamefont {J.~F.}\ \bibnamefont {Douglas}}, \bibinfo
  {author} {\bibfnamefont {J.}~\bibnamefont {Pyun}}, \ and\ \bibinfo {author}
  {\bibfnamefont {A.}~\bibnamefont {Karim}},\ }\href@noop {} {\bibfield
  {journal} {\bibinfo  {journal} {Journal of Polymer Science Part B: Polymer
  Physics}\ }\textbf {\bibinfo {volume} {46}},\ \bibinfo {pages} {2267}
  (\bibinfo {year} {2008})}\BibitemShut {NoStop}%
\bibitem [{\citenamefont {Kikura}\ \emph {et~al.}(2007)\citenamefont {Kikura},
  \citenamefont {Matsushita}, \citenamefont {Kakuta}, \citenamefont {Aritomi},\
  and\ \citenamefont {Kobayashi}}]{kikura2007cluster}%
  \BibitemOpen
  \bibfield  {author} {\bibinfo {author} {\bibfnamefont {H.}~\bibnamefont
  {Kikura}}, \bibinfo {author} {\bibfnamefont {J.}~\bibnamefont {Matsushita}},
  \bibinfo {author} {\bibfnamefont {N.}~\bibnamefont {Kakuta}}, \bibinfo
  {author} {\bibfnamefont {M.}~\bibnamefont {Aritomi}}, \ and\ \bibinfo
  {author} {\bibfnamefont {Y.}~\bibnamefont {Kobayashi}},\ }\href@noop {}
  {\bibfield  {journal} {\bibinfo  {journal} {Journal of materials processing
  technology}\ }\textbf {\bibinfo {volume} {181}},\ \bibinfo {pages} {93}
  (\bibinfo {year} {2007})}\BibitemShut {NoStop}%
\bibitem [{\citenamefont {Gefen}\ \emph {et~al.}(1980)\citenamefont {Gefen},
  \citenamefont {Mandelbrot},\ and\ \citenamefont
  {Aharony}}]{gefen1980critical}%
  \BibitemOpen
  \bibfield  {author} {\bibinfo {author} {\bibfnamefont {Y.}~\bibnamefont
  {Gefen}}, \bibinfo {author} {\bibfnamefont {B.~B.}\ \bibnamefont
  {Mandelbrot}}, \ and\ \bibinfo {author} {\bibfnamefont {A.}~\bibnamefont
  {Aharony}},\ }\href@noop {} {\bibfield  {journal} {\bibinfo  {journal}
  {Physical Review Letters}\ }\textbf {\bibinfo {volume} {45}},\ \bibinfo
  {pages} {855} (\bibinfo {year} {1980})}\BibitemShut {NoStop}%
\bibitem [{\citenamefont {Koza}\ and\ \citenamefont
  {Ausloos}(2007)}]{koza2007ising}%
  \BibitemOpen
  \bibfield  {author} {\bibinfo {author} {\bibfnamefont {Z.}~\bibnamefont
  {Koza}}\ and\ \bibinfo {author} {\bibfnamefont {M.}~\bibnamefont {Ausloos}},\
  }\href@noop {} {\bibfield  {journal} {\bibinfo  {journal} {Physica A:
  Statistical Mechanics and its Applications}\ }\textbf {\bibinfo {volume}
  {375}},\ \bibinfo {pages} {199} (\bibinfo {year} {2007})}\BibitemShut
  {NoStop}%
\bibitem [{\citenamefont {Najafi}(2016{\natexlab{a}})}]{najafi2016monte}%
  \BibitemOpen
  \bibfield  {author} {\bibinfo {author} {\bibfnamefont {M.}~\bibnamefont
  {Najafi}},\ }\href@noop {} {\bibfield  {journal} {\bibinfo  {journal}
  {Physics Letters A}\ }\textbf {\bibinfo {volume} {380}},\ \bibinfo {pages}
  {370} (\bibinfo {year} {2016}{\natexlab{a}})}\BibitemShut {NoStop}%
\bibitem [{\citenamefont {Cambier}\ and\ \citenamefont
  {Nauenberg}(1986)}]{cambier1986distribution}%
  \BibitemOpen
  \bibfield  {author} {\bibinfo {author} {\bibfnamefont {J.}~\bibnamefont
  {Cambier}}\ and\ \bibinfo {author} {\bibfnamefont {M.}~\bibnamefont
  {Nauenberg}},\ }\href@noop {} {\bibfield  {journal} {\bibinfo  {journal}
  {Physical Review B}\ }\textbf {\bibinfo {volume} {34}},\ \bibinfo {pages}
  {8071} (\bibinfo {year} {1986})}\BibitemShut {NoStop}%
\bibitem [{\citenamefont {Coniglio}(1989)}]{coniglio1989fractal}%
  \BibitemOpen
  \bibfield  {author} {\bibinfo {author} {\bibfnamefont {A.}~\bibnamefont
  {Coniglio}},\ }\href@noop {} {\bibfield  {journal} {\bibinfo  {journal}
  {Physical review letters}\ }\textbf {\bibinfo {volume} {62}},\ \bibinfo
  {pages} {3054} (\bibinfo {year} {1989})}\BibitemShut {NoStop}%
\bibitem [{\citenamefont {Coniglio}\ and\ \citenamefont
  {Klein}(1980)}]{coniglio1980clusters}%
  \BibitemOpen
  \bibfield  {author} {\bibinfo {author} {\bibfnamefont {A.}~\bibnamefont
  {Coniglio}}\ and\ \bibinfo {author} {\bibfnamefont {W.}~\bibnamefont
  {Klein}},\ }\href@noop {} {\bibfield  {journal} {\bibinfo  {journal} {Journal
  of Physics A: Mathematical and General}\ }\textbf {\bibinfo {volume} {13}},\
  \bibinfo {pages} {2775} (\bibinfo {year} {1980})}\BibitemShut {NoStop}%
\bibitem [{\citenamefont {Wang}\ and\ \citenamefont
  {Stauffer}(1990)}]{wang1990fractal}%
  \BibitemOpen
  \bibfield  {author} {\bibinfo {author} {\bibfnamefont {J.-S.}\ \bibnamefont
  {Wang}}\ and\ \bibinfo {author} {\bibfnamefont {D.}~\bibnamefont
  {Stauffer}},\ }\href@noop {} {\bibfield  {journal} {\bibinfo  {journal}
  {Zeitschrift f{\"u}r Physik B Condensed Matter}\ }\textbf {\bibinfo {volume}
  {78}},\ \bibinfo {pages} {145} (\bibinfo {year} {1990})}\BibitemShut
  {NoStop}%
\bibitem [{\citenamefont {Saberi}(2009)}]{saberi2009thermal}%
  \BibitemOpen
  \bibfield  {author} {\bibinfo {author} {\bibfnamefont {A.~A.}\ \bibnamefont
  {Saberi}},\ }\href@noop {} {\bibfield  {journal} {\bibinfo  {journal}
  {Journal of Statistical Mechanics: Theory and Experiment}\ }\textbf {\bibinfo
  {volume} {2009}},\ \bibinfo {pages} {P07030} (\bibinfo {year}
  {2009})}\BibitemShut {NoStop}%
\bibitem [{\citenamefont {Davatolhagh}\ \emph {et~al.}(2012)\citenamefont
  {Davatolhagh}, \citenamefont {Moshfeghian},\ and\ \citenamefont
  {Saberi}}]{davatolhagh2012critical}%
  \BibitemOpen
  \bibfield  {author} {\bibinfo {author} {\bibfnamefont {S.}~\bibnamefont
  {Davatolhagh}}, \bibinfo {author} {\bibfnamefont {M.}~\bibnamefont
  {Moshfeghian}}, \ and\ \bibinfo {author} {\bibfnamefont {A.}~\bibnamefont
  {Saberi}},\ }\href@noop {} {\bibfield  {journal} {\bibinfo  {journal}
  {Journal of Statistical Mechanics: Theory and Experiment}\ }\textbf {\bibinfo
  {volume} {2012}},\ \bibinfo {pages} {P02015} (\bibinfo {year}
  {2012})}\BibitemShut {NoStop}%
\bibitem [{\citenamefont {Scholten}\ and\ \citenamefont
  {Kaufman}(1997)}]{scholten1997monte}%
  \BibitemOpen
  \bibfield  {author} {\bibinfo {author} {\bibfnamefont {P.}~\bibnamefont
  {Scholten}}\ and\ \bibinfo {author} {\bibfnamefont {M.}~\bibnamefont
  {Kaufman}},\ }\href@noop {} {\bibfield  {journal} {\bibinfo  {journal}
  {Physical Review B}\ }\textbf {\bibinfo {volume} {56}},\ \bibinfo {pages}
  {59} (\bibinfo {year} {1997})}\BibitemShut {NoStop}%
\bibitem [{\citenamefont {Wilkinson}\ and\ \citenamefont
  {Willemsen}(1983)}]{wilkinson1983invasion}%
  \BibitemOpen
  \bibfield  {author} {\bibinfo {author} {\bibfnamefont {D.}~\bibnamefont
  {Wilkinson}}\ and\ \bibinfo {author} {\bibfnamefont {J.~F.}\ \bibnamefont
  {Willemsen}},\ }\href@noop {} {\bibfield  {journal} {\bibinfo  {journal}
  {Journal of Physics A: Mathematical and General}\ }\textbf {\bibinfo {volume}
  {16}},\ \bibinfo {pages} {3365} (\bibinfo {year} {1983})}\BibitemShut
  {NoStop}%
\bibitem [{\citenamefont {Najafi}\ and\ \citenamefont
  {Ghaedi}(2015)}]{najafi2015geometrical}%
  \BibitemOpen
  \bibfield  {author} {\bibinfo {author} {\bibfnamefont {M.}~\bibnamefont
  {Najafi}}\ and\ \bibinfo {author} {\bibfnamefont {M.}~\bibnamefont
  {Ghaedi}},\ }\href@noop {} {\bibfield  {journal} {\bibinfo  {journal}
  {Physica A: Statistical Mechanics and its Applications}\ }\textbf {\bibinfo
  {volume} {427}},\ \bibinfo {pages} {82} (\bibinfo {year} {2015})}\BibitemShut
  {NoStop}%
\bibitem [{\citenamefont {Najafi}\ \emph {et~al.}(2016)\citenamefont {Najafi},
  \citenamefont {Ghaedi},\ and\ \citenamefont
  {Moghimi-Araghi}}]{najafi2016water}%
  \BibitemOpen
  \bibfield  {author} {\bibinfo {author} {\bibfnamefont {M.}~\bibnamefont
  {Najafi}}, \bibinfo {author} {\bibfnamefont {M.}~\bibnamefont {Ghaedi}}, \
  and\ \bibinfo {author} {\bibfnamefont {S.}~\bibnamefont {Moghimi-Araghi}},\
  }\href@noop {} {\bibfield  {journal} {\bibinfo  {journal} {Physica A:
  Statistical Mechanics and its Applications}\ }\textbf {\bibinfo {volume}
  {445}},\ \bibinfo {pages} {102} (\bibinfo {year} {2016})}\BibitemShut
  {NoStop}%
\bibitem [{\citenamefont {Najafi}(2016{\natexlab{b}})}]{najafi2016bak}%
  \BibitemOpen
  \bibfield  {author} {\bibinfo {author} {\bibfnamefont {M.}~\bibnamefont
  {Najafi}},\ }\href@noop {} {\bibfield  {journal} {\bibinfo  {journal}
  {Journal of Physics A: Mathematical and Theoretical}\ }\textbf {\bibinfo
  {volume} {49}},\ \bibinfo {pages} {335003} (\bibinfo {year}
  {2016}{\natexlab{b}})}\BibitemShut {NoStop}%
\bibitem [{\citenamefont {Dhar}\ and\ \citenamefont
  {Majumdar}(1990)}]{dhar1990abelian}%
  \BibitemOpen
  \bibfield  {author} {\bibinfo {author} {\bibfnamefont {D.}~\bibnamefont
  {Dhar}}\ and\ \bibinfo {author} {\bibfnamefont {S.}~\bibnamefont
  {Majumdar}},\ }\href@noop {} {\bibfield  {journal} {\bibinfo  {journal}
  {Journal of Physics A: Mathematical and General}\ }\textbf {\bibinfo {volume}
  {23}},\ \bibinfo {pages} {4333} (\bibinfo {year} {1990})}\BibitemShut
  {NoStop}%
\bibitem [{\citenamefont {Daerden}\ \emph {et~al.}(2001)\citenamefont
  {Daerden}, \citenamefont {Priezzhev},\ and\ \citenamefont
  {Vanderzande}}]{daerden2001waves}%
  \BibitemOpen
  \bibfield  {author} {\bibinfo {author} {\bibfnamefont {F.}~\bibnamefont
  {Daerden}}, \bibinfo {author} {\bibfnamefont {V.~B.}\ \bibnamefont
  {Priezzhev}}, \ and\ \bibinfo {author} {\bibfnamefont {C.}~\bibnamefont
  {Vanderzande}},\ }\href@noop {} {\bibfield  {journal} {\bibinfo  {journal}
  {Physica A: Statistical Mechanics and its Applications}\ }\textbf {\bibinfo
  {volume} {292}},\ \bibinfo {pages} {43} (\bibinfo {year} {2001})}\BibitemShut
  {NoStop}%
\bibitem [{\citenamefont {Bak}\ \emph {et~al.}(1987)\citenamefont {Bak},
  \citenamefont {Tang},\ and\ \citenamefont {Wiesenfeld}}]{Bak1987Self}%
  \BibitemOpen
  \bibfield  {author} {\bibinfo {author} {\bibfnamefont {P.}~\bibnamefont
  {Bak}}, \bibinfo {author} {\bibfnamefont {C.}~\bibnamefont {Tang}}, \ and\
  \bibinfo {author} {\bibfnamefont {K.}~\bibnamefont {Wiesenfeld}},\ }\href
  {\doibase 10.1103/PhysRevLett.59.381} {\bibfield  {journal} {\bibinfo
  {journal} {Phys. Rev. Lett.}\ }\textbf {\bibinfo {volume} {59}},\ \bibinfo
  {pages} {381} (\bibinfo {year} {1987})}\BibitemShut {NoStop}%
\bibitem [{\citenamefont {Dhar}(1990)}]{Dhar1990Self}%
  \BibitemOpen
  \bibfield  {author} {\bibinfo {author} {\bibfnamefont {D.}~\bibnamefont
  {Dhar}},\ }\href {\doibase 10.1103/PhysRevLett.64.1613} {\bibfield  {journal}
  {\bibinfo  {journal} {Phys. Rev. Lett.}\ }\textbf {\bibinfo {volume} {64}},\
  \bibinfo {pages} {1613} (\bibinfo {year} {1990})}\BibitemShut {NoStop}%
\bibitem [{\citenamefont {Dhar}(1999)}]{Dhar1999Abelian}%
  \BibitemOpen
  \bibfield  {author} {\bibinfo {author} {\bibfnamefont {D.}~\bibnamefont
  {Dhar}},\ }\href {\doibase 10.1016/S0378-4371(98)00493-2} {\bibfield
  {journal} {\bibinfo  {journal} {Physica A: Statistical Mechanics and its
  Applications}\ }\textbf {\bibinfo {volume} {263}},\ \bibinfo {pages} {4}
  (\bibinfo {year} {1999})}\BibitemShut {NoStop}%
\bibitem [{\citenamefont {Majumdar}\ and\ \citenamefont
  {Dhar}(1992)}]{majumdar1992equivalence}%
  \BibitemOpen
  \bibfield  {author} {\bibinfo {author} {\bibfnamefont {S.}~\bibnamefont
  {Majumdar}}\ and\ \bibinfo {author} {\bibfnamefont {D.}~\bibnamefont
  {Dhar}},\ }\href {\doibase http://dx.doi.org/10.1016/0378-4371(92)90447-X}
  {\bibfield  {journal} {\bibinfo  {journal} {Physica A: Statistical Mechanics
  and its Applications}\ }\textbf {\bibinfo {volume} {185}},\ \bibinfo {pages}
  {129 } (\bibinfo {year} {1992})}\BibitemShut {NoStop}%
\bibitem [{\citenamefont {Najafi}(2013{\natexlab{a}})}]{najafi2013left}%
  \BibitemOpen
  \bibfield  {author} {\bibinfo {author} {\bibfnamefont {M.}~\bibnamefont
  {Najafi}},\ }\href@noop {} {\bibfield  {journal} {\bibinfo  {journal}
  {Physical Review E}\ }\textbf {\bibinfo {volume} {87}},\ \bibinfo {pages}
  {062105} (\bibinfo {year} {2013}{\natexlab{a}})}\BibitemShut {NoStop}%
\bibitem [{\citenamefont {Azimi-Tafreshi}\ \emph {et~al.}(2011)\citenamefont
  {Azimi-Tafreshi}, \citenamefont {Lotfi},\ and\ \citenamefont
  {Moghimi-Araghi}}]{Azimi2011Continuous}%
  \BibitemOpen
  \bibfield  {author} {\bibinfo {author} {\bibfnamefont {N.}~\bibnamefont
  {Azimi-Tafreshi}}, \bibinfo {author} {\bibfnamefont {E.}~\bibnamefont
  {Lotfi}}, \ and\ \bibinfo {author} {\bibfnamefont {S.}~\bibnamefont
  {Moghimi-Araghi}},\ }\href {\doibase 10.1142/S0217979211052654} {\bibfield
  {journal} {\bibinfo  {journal} {International Journal of Modern Physics B}\
  }\textbf {\bibinfo {volume} {25}},\ \bibinfo {pages} {4709} (\bibinfo {year}
  {2011})}\BibitemShut {NoStop}%
\bibitem [{\citenamefont {Majumdar}\ and\ \citenamefont
  {Dhar}(1991)}]{Majumdar1991Height}%
  \BibitemOpen
  \bibfield  {author} {\bibinfo {author} {\bibfnamefont {S.~N.}\ \bibnamefont
  {Majumdar}}\ and\ \bibinfo {author} {\bibfnamefont {D.}~\bibnamefont
  {Dhar}},\ }\href {http://stacks.iop.org/0305-4470/24/i=7/a=008} {\bibfield
  {journal} {\bibinfo  {journal} {Journal of Physics A: Mathematical and
  General}\ }\textbf {\bibinfo {volume} {24}},\ \bibinfo {pages} {L357}
  (\bibinfo {year} {1991})}\BibitemShut {NoStop}%
\bibitem [{\citenamefont {Majumdar}(1992)}]{Majumdar1992Exact}%
  \BibitemOpen
  \bibfield  {author} {\bibinfo {author} {\bibfnamefont {S.~N.}\ \bibnamefont
  {Majumdar}},\ }\href {\doibase 10.1103/PhysRevLett.68.2329} {\bibfield
  {journal} {\bibinfo  {journal} {Phys. Rev. Lett.}\ }\textbf {\bibinfo
  {volume} {68}},\ \bibinfo {pages} {2329} (\bibinfo {year}
  {1992})}\BibitemShut {NoStop}%
\bibitem [{\citenamefont {Manna}\ \emph {et~al.}(1990)\citenamefont {Manna},
  \citenamefont {Kiss},\ and\ \citenamefont {Kert{\'e}sz}}]{Manna1990Cascades}%
  \BibitemOpen
  \bibfield  {author} {\bibinfo {author} {\bibfnamefont {S.}~\bibnamefont
  {Manna}}, \bibinfo {author} {\bibfnamefont {L.~B.}\ \bibnamefont {Kiss}}, \
  and\ \bibinfo {author} {\bibfnamefont {J.}~\bibnamefont {Kert{\'e}sz}},\
  }\href {\doibase 10.1007/BF01027312} {\bibfield  {journal} {\bibinfo
  {journal} {Journal of statistical physics}\ }\textbf {\bibinfo {volume}
  {61}},\ \bibinfo {pages} {923} (\bibinfo {year} {1990})}\BibitemShut
  {NoStop}%
\bibitem [{\citenamefont {L{\"u}beck}\ and\ \citenamefont
  {Usadel}(1997{\natexlab{a}})}]{Lubeck1997BTW}%
  \BibitemOpen
  \bibfield  {author} {\bibinfo {author} {\bibfnamefont {S.}~\bibnamefont
  {L{\"u}beck}}\ and\ \bibinfo {author} {\bibfnamefont {K.~D.}\ \bibnamefont
  {Usadel}},\ }\href {\doibase 10.1103/PhysRevE.56.5138} {\bibfield  {journal}
  {\bibinfo  {journal} {Phys. Rev. E}\ }\textbf {\bibinfo {volume} {56}},\
  \bibinfo {pages} {5138} (\bibinfo {year} {1997}{\natexlab{a}})}\BibitemShut
  {NoStop}%
\bibitem [{\citenamefont {Ktitarev}\ \emph {et~al.}(2000)\citenamefont
  {Ktitarev}, \citenamefont {L\"ubeck}, \citenamefont {Grassberger},\ and\
  \citenamefont {B.~Priezzhev}}]{Ktitarev2000Scaling}%
  \BibitemOpen
  \bibfield  {author} {\bibinfo {author} {\bibfnamefont {D.~V.}\ \bibnamefont
  {Ktitarev}}, \bibinfo {author} {\bibfnamefont {S.}~\bibnamefont {L\"ubeck}},
  \bibinfo {author} {\bibfnamefont {P.}~\bibnamefont {Grassberger}}, \ and\
  \bibinfo {author} {\bibfnamefont {V.}~\bibnamefont {B.~Priezzhev}},\ }\href
  {\doibase 10.1103/PhysRevE.61.81} {\bibfield  {journal} {\bibinfo  {journal}
  {Phys. Rev. E}\ }\textbf {\bibinfo {volume} {61}},\ \bibinfo {pages} {81}
  (\bibinfo {year} {2000})}\BibitemShut {NoStop}%
\bibitem [{\citenamefont {Najafi}(2013{\natexlab{b}})}]{najafi2013numerical}%
  \BibitemOpen
  \bibfield  {author} {\bibinfo {author} {\bibfnamefont {M.}~\bibnamefont
  {Najafi}},\ }\href@noop {} {\bibfield  {journal} {\bibinfo  {journal}
  {Physica A: Statistical Mechanics and its Applications}\ }\textbf {\bibinfo
  {volume} {392}},\ \bibinfo {pages} {5179} (\bibinfo {year}
  {2013}{\natexlab{b}})}\BibitemShut {NoStop}%
\bibitem [{\citenamefont {Asasi}\ \emph {et~al.}(2015)\citenamefont {Asasi},
  \citenamefont {Moghimi-Araghi},\ and\ \citenamefont
  {Najafi}}]{asasi2015continuous}%
  \BibitemOpen
  \bibfield  {author} {\bibinfo {author} {\bibfnamefont {H.}~\bibnamefont
  {Asasi}}, \bibinfo {author} {\bibfnamefont {S.}~\bibnamefont
  {Moghimi-Araghi}}, \ and\ \bibinfo {author} {\bibfnamefont {M.}~\bibnamefont
  {Najafi}},\ }\href@noop {} {\bibfield  {journal} {\bibinfo  {journal}
  {Physica A: Statistical Mechanics and its Applications}\ }\textbf {\bibinfo
  {volume} {419}},\ \bibinfo {pages} {196} (\bibinfo {year}
  {2015})}\BibitemShut {NoStop}%
\bibitem [{\citenamefont {L{\"u}beck}\ and\ \citenamefont
  {Usadel}(1997{\natexlab{b}})}]{Lubeck1997Numerical}%
  \BibitemOpen
  \bibfield  {author} {\bibinfo {author} {\bibfnamefont {S.}~\bibnamefont
  {L{\"u}beck}}\ and\ \bibinfo {author} {\bibfnamefont {K.~D.}\ \bibnamefont
  {Usadel}},\ }\href {\doibase 10.1103/PhysRevE.55.4095} {\bibfield  {journal}
  {\bibinfo  {journal} {Phys. Rev. E}\ }\textbf {\bibinfo {volume} {55}},\
  \bibinfo {pages} {4095} (\bibinfo {year} {1997}{\natexlab{b}})}\BibitemShut
  {NoStop}%
\bibitem [{\citenamefont {Moghimi-Araghi}\ \emph {et~al.}(2005)\citenamefont
  {Moghimi-Araghi}, \citenamefont {Rajabpour},\ and\ \citenamefont
  {Rouhani}}]{Moghimi2005Abelian}%
  \BibitemOpen
  \bibfield  {author} {\bibinfo {author} {\bibfnamefont {S.}~\bibnamefont
  {Moghimi-Araghi}}, \bibinfo {author} {\bibfnamefont {M.}~\bibnamefont
  {Rajabpour}}, \ and\ \bibinfo {author} {\bibfnamefont {S.}~\bibnamefont
  {Rouhani}},\ }\href {\doibase 10.1016/j.nuclphysb.2005.04.002} {\bibfield
  {journal} {\bibinfo  {journal} {Nuclear Physics B}\ }\textbf {\bibinfo
  {volume} {718}},\ \bibinfo {pages} {362} (\bibinfo {year} {2005})},\ \Eprint
  {http://arxiv.org/abs/0410434} {arXiv:0410434 [cond-mat]} \BibitemShut
  {NoStop}%
\bibitem [{\citenamefont {Najafi}\ \emph
  {et~al.}(2012{\natexlab{a}})\citenamefont {Najafi}, \citenamefont
  {Moghimi-Araghi},\ and\ \citenamefont {Rouhani}}]{najafi2012avalanche}%
  \BibitemOpen
  \bibfield  {author} {\bibinfo {author} {\bibfnamefont {M.}~\bibnamefont
  {Najafi}}, \bibinfo {author} {\bibfnamefont {S.}~\bibnamefont
  {Moghimi-Araghi}}, \ and\ \bibinfo {author} {\bibfnamefont {S.}~\bibnamefont
  {Rouhani}},\ }\href@noop {} {\bibfield  {journal} {\bibinfo  {journal}
  {Physical Review E}\ }\textbf {\bibinfo {volume} {85}},\ \bibinfo {pages}
  {051104} (\bibinfo {year} {2012}{\natexlab{a}})}\BibitemShut {NoStop}%
\bibitem [{\citenamefont {Najafi}(2014)}]{najafi2014bak}%
  \BibitemOpen
  \bibfield  {author} {\bibinfo {author} {\bibfnamefont {M.}~\bibnamefont
  {Najafi}},\ }\href@noop {} {\bibfield  {journal} {\bibinfo  {journal}
  {Physics Letters A}\ }\textbf {\bibinfo {volume} {378}},\ \bibinfo {pages}
  {2008} (\bibinfo {year} {2014})}\BibitemShut {NoStop}%
\bibitem [{\citenamefont {Najafi}\ \emph
  {et~al.}(2012{\natexlab{b}})\citenamefont {Najafi}, \citenamefont
  {Moghimi-Araghi},\ and\ \citenamefont {Rouhani}}]{najafi2012observation}%
  \BibitemOpen
  \bibfield  {author} {\bibinfo {author} {\bibfnamefont {M.}~\bibnamefont
  {Najafi}}, \bibinfo {author} {\bibfnamefont {S.}~\bibnamefont
  {Moghimi-Araghi}}, \ and\ \bibinfo {author} {\bibfnamefont {S.}~\bibnamefont
  {Rouhani}},\ }\href@noop {} {\bibfield  {journal} {\bibinfo  {journal}
  {Journal of Physics A: Mathematical and Theoretical}\ }\textbf {\bibinfo
  {volume} {45}},\ \bibinfo {pages} {095001} (\bibinfo {year}
  {2012}{\natexlab{b}})}\BibitemShut {NoStop}%
\bibitem [{\citenamefont {Dashti-Naserabadi}\ and\ \citenamefont
  {Najafi}(2015)}]{dashti2015statistical}%
  \BibitemOpen
  \bibfield  {author} {\bibinfo {author} {\bibfnamefont {H.}~\bibnamefont
  {Dashti-Naserabadi}}\ and\ \bibinfo {author} {\bibfnamefont {M.}~\bibnamefont
  {Najafi}},\ }\href@noop {} {\bibfield  {journal} {\bibinfo  {journal}
  {Physical Review E}\ }\textbf {\bibinfo {volume} {91}},\ \bibinfo {pages}
  {052145} (\bibinfo {year} {2015})}\BibitemShut {NoStop}%
\bibitem [{\citenamefont {Dashti-Naserabadi}\ and\ \citenamefont
  {Najafi}(2017)}]{dashti2017bak}%
  \BibitemOpen
  \bibfield  {author} {\bibinfo {author} {\bibfnamefont {H.}~\bibnamefont
  {Dashti-Naserabadi}}\ and\ \bibinfo {author} {\bibfnamefont {M.}~\bibnamefont
  {Najafi}},\ }\href@noop {} {\bibfield  {journal} {\bibinfo  {journal}
  {Physical Review E}\ }\textbf {\bibinfo {volume} {96}},\ \bibinfo {pages}
  {042115} (\bibinfo {year} {2017})}\BibitemShut {NoStop}%
\bibitem [{\citenamefont {Najafi}(2015)}]{najafi2015observation}%
  \BibitemOpen
  \bibfield  {author} {\bibinfo {author} {\bibfnamefont {M.}~\bibnamefont
  {Najafi}},\ }\href@noop {} {\bibfield  {journal} {\bibinfo  {journal}
  {Journal of Statistical Mechanics: Theory and Experiment}\ }\textbf {\bibinfo
  {volume} {2015}},\ \bibinfo {pages} {P05009} (\bibinfo {year}
  {2015})}\BibitemShut {NoStop}%
\bibitem [{\citenamefont {Vespignani}\ \emph {et~al.}(1998)\citenamefont
  {Vespignani}, \citenamefont {Dickman}, \citenamefont {Mu{\~n}oz},\ and\
  \citenamefont {Zapperi}}]{vespignani1998driving}%
  \BibitemOpen
  \bibfield  {author} {\bibinfo {author} {\bibfnamefont {A.}~\bibnamefont
  {Vespignani}}, \bibinfo {author} {\bibfnamefont {R.}~\bibnamefont {Dickman}},
  \bibinfo {author} {\bibfnamefont {M.~A.}\ \bibnamefont {Mu{\~n}oz}}, \ and\
  \bibinfo {author} {\bibfnamefont {S.}~\bibnamefont {Zapperi}},\ }\href@noop
  {} {\bibfield  {journal} {\bibinfo  {journal} {Physical review letters}\
  }\textbf {\bibinfo {volume} {81}},\ \bibinfo {pages} {5676} (\bibinfo {year}
  {1998})}\BibitemShut {NoStop}%
\bibitem [{\citenamefont {Rossi}\ \emph {et~al.}(2000)\citenamefont {Rossi},
  \citenamefont {Pastor-Satorras},\ and\ \citenamefont
  {Vespignani}}]{rossi2000universality}%
  \BibitemOpen
  \bibfield  {author} {\bibinfo {author} {\bibfnamefont {M.}~\bibnamefont
  {Rossi}}, \bibinfo {author} {\bibfnamefont {R.}~\bibnamefont
  {Pastor-Satorras}}, \ and\ \bibinfo {author} {\bibfnamefont {A.}~\bibnamefont
  {Vespignani}},\ }\href@noop {} {\bibfield  {journal} {\bibinfo  {journal}
  {Physical review letters}\ }\textbf {\bibinfo {volume} {85}},\ \bibinfo
  {pages} {1803} (\bibinfo {year} {2000})}\BibitemShut {NoStop}%
\bibitem [{\citenamefont {Karmakar}\ and\ \citenamefont
  {Manna}(2004)}]{karmakar2004directed}%
  \BibitemOpen
  \bibfield  {author} {\bibinfo {author} {\bibfnamefont {R.}~\bibnamefont
  {Karmakar}}\ and\ \bibinfo {author} {\bibfnamefont {S.}~\bibnamefont
  {Manna}},\ }\href@noop {} {\bibfield  {journal} {\bibinfo  {journal}
  {Physical Review E}\ }\textbf {\bibinfo {volume} {69}},\ \bibinfo {pages}
  {067107} (\bibinfo {year} {2004})}\BibitemShut {NoStop}%
\bibitem [{\citenamefont {Najafi}\ and\ \citenamefont
  {Nezhadhaghighi}(2017)}]{najafi2016scale}%
  \BibitemOpen
  \bibfield  {author} {\bibinfo {author} {\bibfnamefont {M.}~\bibnamefont
  {Najafi}}\ and\ \bibinfo {author} {\bibfnamefont {M.~G.}\ \bibnamefont
  {Nezhadhaghighi}},\ }\href@noop {} {\bibfield  {journal} {\bibinfo  {journal}
  {Physical Review E}\ }\textbf {\bibinfo {volume} {95}},\ \bibinfo {pages}
  {032112} (\bibinfo {year} {2017})}\BibitemShut {NoStop}%
\bibitem [{\citenamefont {Tebaldi}\ \emph {et~al.}(1999)\citenamefont
  {Tebaldi}, \citenamefont {De~Menech},\ and\ \citenamefont
  {Stella}}]{tebaldi1999multifractal}%
  \BibitemOpen
  \bibfield  {author} {\bibinfo {author} {\bibfnamefont {C.}~\bibnamefont
  {Tebaldi}}, \bibinfo {author} {\bibfnamefont {M.}~\bibnamefont {De~Menech}},
  \ and\ \bibinfo {author} {\bibfnamefont {A.~L.}\ \bibnamefont {Stella}},\
  }\href@noop {} {\bibfield  {journal} {\bibinfo  {journal} {Physical Review
  Letters}\ }\textbf {\bibinfo {volume} {83}},\ \bibinfo {pages} {3952}
  (\bibinfo {year} {1999})}\BibitemShut {NoStop}%
\bibitem [{\citenamefont {Isichenko}(1992)}]{isichenko1992percolation}%
  \BibitemOpen
  \bibfield  {author} {\bibinfo {author} {\bibfnamefont {M.~B.}\ \bibnamefont
  {Isichenko}},\ }\href@noop {} {\bibfield  {journal} {\bibinfo  {journal}
  {Reviews of modern physics}\ }\textbf {\bibinfo {volume} {64}},\ \bibinfo
  {pages} {961} (\bibinfo {year} {1992})}\BibitemShut {NoStop}%
\end{thebibliography}%

\end{document}